\documentclass[twocolumn,showpacs,showkeys,preprintnumbers,amsmath,amssymb]{revtex4}
\usepackage{amsthm}        \usepackage{dcolumn}        \usepackage{bm}
\usepackage{graphicx} \usepackage{subfigure}

\newcommand{\N}{\mathbb{N}}                 \newcommand{\Z}{\mathbb{Z}}
\newcommand{\R}{\mathbb{R}}                 \newcommand{\C}{\mathbb{C}}

\newtheorem{claim}{Claim} \newtheorem*{cor}{Corollary}
\begin{document}

\title{On the quantum spectrum of isochronous potentials}

\author{J. Dorignac}

\affiliation{College  of  Engineering,  Boston University,  Boston  MA
02215}

\date{\today}
\begin{abstract}
In  this paper, the quantum spectrum of isochronous potentials is 
investigated.  Given that the frequency  of the classical  motion in
such potentials  is energy-independent, it is natural  to expect their
quantum spectra  to  be  equispaced. However, as it has already been 
shown in some specific examples, this property is not always true. 
To gain some general insight into this problem, 
a WKB analysis of the spectrum, 
valid for any analytic potential, is
performed and the first
semiclassical corrections to its regular spacing are calculated. 
We illustrate the results on the two-parameter
family  of isochronous potentials  derived in  \cite{Still89},  which
includes the  harmonic oscillator, the asymmetric parabolic well, the
radial harmonic  oscillator and Urabe's potential  as special limiting
cases. In addition, some new analytical expressions for
families of isochronous potentials and their corresponding spectra are 
derived by means of the above-mentioned method.
\end{abstract}

\pacs{45.05.+x; 03.65.Sq}
\keywords{Isochronous; WKB method; Semiclassical spectrum}

\maketitle

\section{Introduction}
Several studies  have already been devoted to  the interesting problem
of   the   relation   between   classical  and   quantum  ``generalised
harmonicity''. In  classical mechanics, the  motion of an  oscillator in
the  parabolic  well $V(x)=\omega^2  x^2/2$  is  known  to possess  an
energy-independent  frequency $\omega$.  However, this  harmonicity is
not specific to the latter and  can  be  achieved by  designing
adequate   potentials  called {\em isochronous}. Their construction 
simply amounts to shearing  the parabolic
well in  such a way  that, at fixed  energy, the distance  between two
turning points is preserved (see for example \cite{Osyp87,Bol02}).

In  quantum mechanics,  the energy  levels of  a parabolic  well are
regularly  spaced   by  a  quantity  $\hbar   \omega$. Thus, a generalised
harmonicity   is  naturally   defined    by   an   equispaced
spectrum.  Similar to the classical  case,  it is possible to
construct potentials,  essentially different from  the parabolic well,
whose spectrum is exactly harmonic.  This can be done by applying 
a supersymmetric transform followed by a Darboux transform to the 
harmonic well which lead to families of potentials called {\em isospectral}  
partners of
the   harmonic   oscillator   (for   a   review,   see   for   example
\cite{Cooper95}).  A  famous example  of  such  a  potential has  been
derived by Abraham and Moses \cite{Abraham80}. 

A link between these classical and quantum transformations has been 
established by Eleonskii et al. \cite{Eleonskii97}. 
They show that the classical limit of 
the isospectral transformation is precisely the isochronism preserving 
shear previously mentioned. 
The next question is 
whether the two classes of classical and quantum generalised harmonic 
potentials  are the  same. Surprisingly  enough, they are not.  
Both aspects of  this  problem have  been investigated  by
several  authors who demonstrated through specific examples that  neither
isochronism generates  an equispaced  spectrum nor does a regularly spaced
spectrum stem from an isochronous potential in general
\cite{NietoGut81,Nieto81,Ghosh81,Still89,Mohaz00}.

Despite this inequivalence, it  is worth noting that the semiclassical
Einstein-Brillouin-Keller   (EBK)  quantisation   of   an  isochronous
potential   leads   to  an   equispaced   EBK-spectrum.  Indeed,   the
semiclassical quantisation rule is  $    I_0(E)   =   \hbar
(n+\frac{1}{2}),\ n  \in \N$ where  the action $I_0(E)$ is  related to
the  frequency   $\omega(E)$  and  the   energy  $E$  by   $dI_0/dE  =
\omega(E)^{-1}$.   As the  frequency  of an  isochronous potential  is
energy-independent, $E = \omega  I_0$ and EBK levels are
given  by $E^{\scriptscriptstyle \rm EBK}_n = 
(n+\frac{1}{2})\hbar  \omega$.  
Thus, their spacing is constant, $E^{\scriptscriptstyle \rm EBK}_{n+1}
-E^{\scriptscriptstyle \rm EBK}_n=\hbar \omega$.

Consequently,  as  the   usual  semiclassical  quantisation  fails  to
distinguish  between the spectrum  of an  isochronous potential  and a
purely harmonic spectrum, studies to date have resorted either to
the  study  of  isochronous  potentials  whose  quantum  spectrum  (or
quantisation condition) is  exactly known or to numerics.  To date, 
the asymmetric parabolic (also called {\em split-harmonic}) well
\footnote{See \cite{Still89}, the latter has the form $V(x)=\omega_1^2
  x^2/2$ for $x<0$ and $V(x)=\omega_2^2 x^2/2$ for $x>0$.} is the only
  isochronous potential whose  exact quantisation condition proves its
  quantum      levels     to      be     non  strictly equidistant
  \cite{Ghosh81,Still89}.  Another class  of potentials  defined  on a
  finite  interval  of the  real  axis,  and  first derived  by  Urabe
  \cite{Urabe61},  have been  investigated  analytically by  Mohazzabi
  \cite{Mohaz00}. But no closed form for the quantisation is available
  in this  case and the  lack of regularity  in the spacings  has been
  computed numerically.

Finally,  Stillinger  and Stillinger  \cite{Still89}  have proposed  a
two-parameter family  of isochronous potentials  interpolating between
the harmonic  oscillator and  the asymmetric parabolic  well. Although
not  explicitly mentioned  in their  paper, this  family  includes the
radial harmonic  potential (also known as the {\em isotonic} potential) and
Urabe's potential. For  non extremal values of the parameters,
this family is analytic on the  entire real line. It also includes the
one-parameter  family of  analytic potentials  defined by  Bolotin and
MacKay  \cite{Bol02}.  To  prove  that  the exact  spectrum  of  these
potentials  is not  equidistant in  general, the  authors  based their
argument on  the assumption that energy levels  depend continuously on
the parameters.  As the  family interpolates continuously  between the
harmonic oscillator,  whose spectrum  is equispaced, and  the asymmetric
parabolic well, whose spectrum is not equispaced, 
they deduced that the spectrum is not  exactly equidistant for some range
of the  parameters. However,
this result is  only qualitative in the sense that  the authors do not
provide a method to evaluate  the spectrum for general values of the
parameters.

The aim  of the present paper is  precisely to fill in this gap and to
provide a way to calculate the first corrections to the equispacing of
the EBK spectrum. In section \ref{Iso}, we first review some
properties of  isochronous potentials and  derive  {\em new
analytical  expressions} for  some one-  or two-parameter  families of
isochronous potentials. We then use  the perturbation method WKB up to
fourth order (beyond the EBK quantisation) to derive an expression for the
corrections  to the  equispacing  valid for  any analytic  isochronous
potential (section \ref{WKB2nd}).   In section \ref{Twoparafamily}, we
apply these  results to the  potentials described above and  derive an
analytical or  asymptotic expression for their  {\em higher-order WKB}
quantisation  condition. The  resulting higher-order  WKB  spectra are
then  checked  against  numerical  evaluations of  the  exact  quantum
problem and  prove not  only to reproduce  the right behaviour  of the
quantum levels  at high energy  but also to  be quite accurate  at low
energy for some  range of the parameters. A discussion of these results
and some conclusions are given in section \ref{Discuss}.

\section{Isochronous potentials} \label{Iso}

\subsection{Generalities} \label{Gener}

The literature  about isochronous potentials (or  more generally about
isochronous centres) is particularly vast and the interested reader is
referred to \cite{Chavarriga99} and references therein for
more information.  This section is  not intended to present a detailed
analysis of isochronism  but merely to recall some  basic facts and to
introduce certain notations  which will be useful in  the rest of this
paper. Our approach of isochronism is mainly inspired by 
\cite{Robnik99} and based on the so-called $S$ {\em function} (see below
for a definition). Another fruitful way to obtain explicit analytical 
expressions for isochronous potentials has been developed in 
\cite{Gonera}.

A potential $V(x)$ is said to  be isochronous if it generates a motion
$x(t)$,   obeying  $\ddot{x}+V'(x)=0$,   whose  period   (frequency)  is
energy-independent. It can be shown that such a potential has a single
minimum (see  for  instance  \cite{Bol02}). For potentials which are at 
least twice differentiable at the origin, we  will consider
$V(0)=0$, $V'(0)=0$  and $V''(0)=\omega^2$ throughout,  without loss of
generality.

For arbitrary potentials, the period of motion is
\begin{equation} \label{periodE}
T(E)        =        \sqrt{2}        \int\limits_{x_{-}(E)}^{x_{+}(E)}
\frac{dx}{\sqrt{E-V(x)}}
\end{equation}  
where $x_{\pm}(E)$  are the  two turning points  of the  trajectory at
energy  $E$. Conversely,  it has  been  shown by  Landau and  Lifshitz
\cite{Landau}  that, for  a  prescribed  period  $T(E)$, there is a  
corresponding 
distance between the turning points given by
\begin{equation} \label{invLand}
x_{+}(E)   -  x_{-}(E)   =   \frac{1}{\pi  \sqrt{2}}   \int\limits_0^E
\frac{T(u)}{\sqrt{E-u}}\, du \ .
\end{equation}  
Imposing isochronism to the  potential ($T(E)= \text{cst}  =
2\pi/\omega$) yields
\begin{equation} \label{Delta}
x_{+}(E) - x_{-}(E) = 2\frac{\sqrt{2E}}{\omega}\ .
\end{equation} 
If we let $\hat{x}(E)$ be   the   middle    of    the   segment
$[x_{-}(E),x_{+}(E)]$, we obtain the equivalent formulation
\begin{equation} \label{xpmE}
x_{\pm}(E) = \pm \frac{\sqrt{2E}}{\omega} + \hat{x}(E)\ .
\end{equation} 

An entire freedom is left in the choice of $\hat{x}(E)$ provided the
inversion  of (\ref{xpmE}) leads  to a  potential $V(x)\,  (\equiv E)$
single-valued on the real  axis. The smoothness of $\hat{x}$ induces
the   smoothness  of  $V$.    Geometrically,  the  function
$\hat{x}(E)$ defines a {\em shear} of the parabolic potential giving
rise to another  isochronous potential (for an illustration  of such a
transformation see \cite{Osyp87}).

For our  purpose, it's convenient to rephrase  the results above  in a
slightly different way. Let us define  a variable $X \in \R$ through the
transformation
\begin{equation} \label{Xdef}
V(x) = \frac{1}{2} \omega^2 X^2, \ \ \frac{dx}{dX} > 0\ .
\end{equation} 
The two eqs. (\ref{xpmE}) can now be recast into a single one
\begin{equation} \label{xofX}
x(X) = X + \bar{x}(X)\ .
\end{equation} 
The function $x(X)$ now represents the two branches  of the potential
$V$ parametrised by  $X$. The new shear function  $\bar{x}$ is related
to $\hat{x}$  by $\bar{x}(X) =  \hat{x}(\frac{1}{2} \omega^2 X^2)$
and is even in $X$,
\begin{equation} \label{xbareven}
\bar{x}(X) = \bar{x}(-X)\ .
\end{equation} 
As    $V(0)=0$, equations (\ref{Xdef}) and (\ref{xofX}) yield
$x(0)=\bar{x}(0)=0$. The condition  $dx/dX> 0$ of (\ref{Xdef}) ensures
that   $X$  and   $x$  are   in   bijection,  i.e.   that  $V(x)$   is
single-valued.   Notice  that   we  could   have  chosen   $dx/dX  <0$
instead. Now, from (\ref{xofX}),
\begin{equation} \label{reldxdXS}
\frac{dx}{dX} = 1 + S(X)
\end{equation}   
where
\begin{equation} \label{defS}
S(X) = \frac{d\bar{x}}{dX}\ .
\end{equation}   
The latter is similar to the function introduced in \cite{Robnik99} but
our definition of  $X$ does not  include the frequency $\omega$.

Then, $dx/dX> 0 \ \Rightarrow S(X) > -1$. Moreover, as $\bar{x}(X)$ is
even, $S(X)$ is odd,
\begin{equation} \label{Sodd}
S(X) = -S(-X)
\end{equation}   
which finally yields,
\begin{equation} \label{Smod}
\forall X \in \R, \ \ |S(X)| < 1 .
\end{equation}  
Remark  that  the function  $S$  defines the  potential  $V$  up to  a
multiplicative constant only. Indeed, from (\ref{reldxdXS}) we have
\begin{equation} \label{relxXS}
x = X + \int\limits_0^X S(u)\, du\, .
\end{equation} 
Inverting this relation gives $X(x)$ and thus, from (\ref{Xdef}),
\begin{equation} \label{relVomegaX}
V(x) = \frac{1}{2} \omega^2 [X(x)]^2 \, .
\end{equation} 
The  multiplicative  constant is  thus  basically  the  square of  the
frequency, which can be chosen arbitrarily.

\subsection{Scaling properties} \label{scalesec}

In  what follows,  we derive  some simple,  though  important, results
regarding the scaling properties of isochronous potentials.
\begin{claim}
Let $V(x)$ be isochronous with frequency $\omega$. Then
\begin{equation} \label{genscaling}
\forall   (\gamma,\beta)   \in  \R^{*2},   \   \   \  \tilde{V}(x)   =
\left(\frac{\gamma}{\beta}\right)^2 V(\beta x)
\end{equation}
is isochronous with frequency $\tilde{\omega} = \gamma \omega$.
\end{claim}
\begin{proof}
Let  $I_0(E)$ be  the  action  associated to  $V(x)$.  Then $I_0(E)  =
\frac{\sqrt{2}}{\pi}\int_{x_{-}(E)}^{x_{+}(E)}   \sqrt{E-V(x)}\,   dx$
where  $V(x_{\pm}(E))=E$.  Let  $\tilde{I}_0(E)$  be  associated  with
$\tilde{V}(x)$. Its turning points at  energy $E$ are related to those
of $V(x)$ by
\begin{equation} \label{scaledturnpoints}
\tilde{x}_{\pm}(E)    =   \frac{1}{\beta}   x_{\pm}\left(\frac{\beta^2
E}{\gamma^2}\right)\, .
\end{equation}
Hence,
\begin{equation} \label{scaledactions}
\tilde{I}_0(E)    =    \frac{\gamma}{\beta^2}   I_0\left(\frac{\beta^2
    E}{\gamma^2}\right)\, .
\end{equation}
So far, this relation is true for all potentials.

For an isochronous potential,  however, $I_0(E)= E/\omega$. Then, from
the previous  relation, $\tilde{I}_0(E)  = E/ (\gamma  \omega)$, which
proves  that  $\tilde{V}(x)$  is  isochronous with  frequency  $\gamma
\omega$.
\end{proof} 

\begin{cor}
If $V(x)$ is isochronous with frequency $\omega$,
\begin{equation} \label{scaledVsamefreq}
 \tilde{V}(x) = \frac{1}{\beta^2} V(\beta x)
\end{equation}
 is isochronous with the {\em same} frequency.
\end{cor}

This shows that, once an  isochronous potential $V(x)$ is known, a 
one-parameter family of isochronous potentials with the same frequency  
may be derived from the above scaling.

\begin{claim}
The functions $\tilde{\bar{x}}(X)$ and $\tilde{S}(X)$ corresponding to
$\tilde{V}(x)$ are related to  $\bar{x}(X)$ and $S(X)$ defining $V(x)$
by
\begin{equation} \label{scaledxbarS}
\tilde{\bar{x}}(X) = \frac{1}{\beta}  \bar{x}(\beta X)\ \ \ \text{and}
\ \ \ \tilde{S}(X) = S(\beta X)\ .
\end{equation}
\end{claim}
\begin{proof}
By   definition   $\tilde{x}(X)=X+\tilde{\bar{x}}(X)$   which   yields
$\tilde{\bar{x}}(X)=  [\tilde{x}(X)+\tilde{x}(-X)]/2$.  Starting  from
$\tilde{V}(\tilde{x}) =  \frac{1}{2} \tilde{\omega}^2 X^2$,  we obtain
$(\gamma/\beta)^2 V(\beta  \tilde{x}) = \frac{1}{2}  \gamma^2 \omega^2
X^2$,  that  is  $V(\beta  \tilde{x})  =  \frac{1}{2}  \omega^2  (\beta
X)^2$.    Hence,   $\tilde{x}(X)    =    \frac{1}{\beta}   x    (\beta
X)$.  Reinstating  this last  expression  in $\tilde{\bar{x}}(X)$  and
using $\bar{x}(X)= [x(X)+x(-X)]/2$, we  get the first of the relations
(\ref{scaledxbarS}).  The  second  is  obvious  due  to  the  relation
$\tilde{S}(X) = d \tilde{\bar{x}}(X)/ dX$.
\end{proof} 

Note that, as expected, $\beta$ is the parameter involved in
the scaling  relations (\ref{scaledxbarS}) while 
$\gamma$ is an overall prefactor responsible for the tuning of the
frequency only.

\begin{claim} \label{asymptbehavV}
Asymptotic     behaviour      of     $\tilde{V}(x)$     defined     by
(\ref{scaledVsamefreq})  with respect to  $\beta$. Provided  $S(X)$ is
defined on the entire real line,
\begin{equation} \label{Vbeta0}
\text{As}\  \beta  \rightarrow  0\,  , \  \  \tilde{V}(x)  \rightarrow
  \frac{1}{2} \left(\frac{\omega}{1+S_0^+{\rm sgn}(x)}\right)^2 x^2
\end{equation} 
\begin{equation} \label{Vbetainf}
\text{As}\ \beta  \rightarrow \infty\, , \  \ \tilde{V}(x) \rightarrow
\frac{1}{2}  \left(\frac{\omega}{1+\langle  S  \rangle  {\rm  sgn}(x)}
\right)^2 x^2
\end{equation} 
where
$$  S_0^+ =  \lim_{X \rightarrow  0^+} S(X)\  \ {\rm  and}\  \ \langle
  S\rangle = \lim_{T \rightarrow  \infty} \frac{1}{T} \int_0^T \! S(u)
  \, du\ .$$
\end{claim}
The function ${\rm sgn}(x)$ is equal  to 1 if $x>0$ and $-1$ if $x<0$.
If  $S(X)$ is  continuous in  $X=0$, then $S(0)=0$.  The first
expression  then reduces  to  the harmonic  oscillator potential.  The
second  one   corresponds  to   a  split  harmonic   oscillator  whose
frequencies are $\omega_{\pm} = \omega/(1 \pm \langle S \rangle)$.

\begin{proof}
Let us rewrite equation (\ref{relxXS}) in the form
$$  x  = X  \left(  1  + \frac{1}{X}  \int\limits_0^X  \!  S(u) \,  du
\right)\,  .$$  Now,  applied  to  the rescaled  potential  for  which
$\tilde{S}(X)=S(\beta X)$ (see \eqref{scaledxbarS}), this gives
$$x = X \left( 1  + \frac{1}{X} \int\limits_0^X \!  \tilde{S}(u) \, du
  \right) = X \left( 1  + \frac{1}{\beta X} \int\limits_0^{\beta X} \!
  S(v) \,  dv \right) $$ Taking  the limits $\beta  \rightarrow 0^+$ and
  $\beta \rightarrow  \infty$ yields  $x =  X \left(  1 +
  S_0^+{\rm sgn}(X) \right)$  and $x = X \left( 1  + \langle S \rangle
  {\rm  sgn}(X)  \right)$ respectively, and, as  
${\rm  sgn}(X)={\rm sgn}(x)$,  we
  obtain (\ref{Vbeta0}) and (\ref{Vbetainf}).
\end{proof}

Now, using the results of the previous proof, 
we can obtain the asymptotic behaviour 
of an isochronous potential defined on the entire real line 
when $|x| \rightarrow \infty$.
\begin{claim} \label{asymptbehavVxinfty} 
Providing  $\left|\langle S \rangle\right| \neq 1$, where $\langle S \rangle$  
is defined in claim \ref{asymptbehavV},
\begin{equation} \label{Vxinfty}
V(x) =
\frac{1}{2}  \left(\frac{\omega}{1+\langle  S  \rangle  {\rm  sgn}(x)}
\right)^2 x^2 +o(x^2),\  |x| \rightarrow \infty
\end{equation}
\end{claim}
When $\left|\langle S \rangle\right| = 1$, the potential is singular. 
In this case, only one branch  (say $x \rightarrow \infty$) 
is asymptotically parabolic and such that 
$V(x) = \frac{\omega^2}{8}x^2+o(x^2)$, $x \rightarrow \infty$. 
\subsection{Examples}

We  now  apply  the  above  considerations  to  the  determination  of
isochronous  potentials  whose  analytical  expression  can  be  given
explicitly.  We start  from  the  function  $S(X)$ which is
required to satisfy  the  conditions  given in (\ref{Sodd})  and
(\ref{Smod}).  We then  apply scaling  (\ref{scaledVsamefreq}) to
derive  a  family  of  potentials  with the  same  frequency.  To  our
knowledge, the families of  potentials derived in section \ref{ssIIC2}
never before appeared in the literature.

\subsubsection{Family I} \label{ssIIC1}

Let
\begin{equation} \label{SStill}
S(X)  = \frac{\alpha  X}{\sqrt{1+\alpha X^2}}\,  ,  \ \  \ \alpha  \in
[0,1]\, .
\end{equation}
The condition on $\alpha$ ensures that $|S(X)| < 1$. Moreover
$S(X)   =  -S(-X)$ so that, conditions
(\ref{Sodd})  and  (\ref{Smod})  are  fulfilled.  The  shear  function
corresponding  to $S(X)$ is  $ \bar{x}(X)  = \int_0^X  \! S(u)\,  du =
\sqrt{1+\alpha X^2} - 1$ and inverting (\ref{relxXS}) leads 
to
$$  X  = \frac{(x+1)-\sqrt{\alpha  x  (x+2)+1}}{1-\alpha}\, .$$  Using
(\ref{relVomegaX}) we obtain the potential
\begin{equation} \label{Vbol}
 V(\alpha;x)  = \frac{\omega^2}{2}  \left(  \frac{(x+1)-\sqrt{\alpha x
 (x+2)+1}}{1-\alpha}\right)^2 \, .
\end{equation}      
This is precisely the {\em one-parameter} family of potentials derived
by  Bolotin and  MacKay  in \cite{Bol02}.  Notice  that the  parameter
$\alpha$ is  distinct from the  scaling parameter $\beta$  introduced in
the previous  section. Thus, relation  (\ref{scaledVsamefreq}) enables
us to derive a {\em two-parameter} family of isochronous potentials
\begin{equation} \label{VStill}
V(\alpha,\beta;x)  =   \frac{\omega^2}{2\beta^2}  \left(  \frac{(\beta
 x+1)-\sqrt{\alpha \beta x (\beta x+2)+1}}{1-\alpha}\right)^2
\end{equation} 
presented for  the first time  by Stillinger et al.  in \cite{Still89}
(cf.   \footnote{This  is  clear   from  the   substitution  $\omega^2
\rightarrow  K$,   $\alpha  \rightarrow  \xi^2$,   $\beta  \rightarrow
\sqrt{\beta}/\xi$}).

The   family  (\ref{VStill})   includes   several  known   isochronous
potentials as limiting cases which are listed below.
\begin{itemize}
\item $\alpha \rightarrow 0$ or $\beta \rightarrow0$: 
{\em Harmonic oscillator}.
\begin{equation} \label{HOsc}
V(x) = \frac{\omega^2}{2} x^2 \, .
\end{equation}
\item $\alpha \rightarrow 1$ and $\beta \neq 0$: {\em Isotonic potential}.
\begin{equation} \label{Isotonic}
 V(x)  =  \frac{\omega^2}{8\beta^2}  \left(  \beta  x+1-\frac{1}{\beta
 x+1}\right)^2 \ \ \ x > -\frac{1}{\beta}
\end{equation}
This  potential,  also  called  radial harmonic  oscillator,  has  the
 property to  be singular  at $x =  -1/\beta$ and  to be defined  on a
 half-line.
\item $\beta \rightarrow \infty$: {\em Split harmonic oscillator}.
\begin{equation} \label{Splitharm}
V(x)        =       \left\{        \begin{array}{cc}       \frac{1}{2}
\left(\frac{\omega}{1+\sqrt{\alpha}}\right)^2 x^2  & ,  \ x \geq  0 \\
\frac{1}{2} \left(\frac{\omega}{1-\sqrt{\alpha}}\right)^2 x^2  & , \ x
\leq 0 \end{array} \right.
\end{equation}
In the case $\alpha \rightarrow 1$,  the left part ($x \leq 0$) of the
potential converges to  a hard wall in $x=0$ and gives  rise to a {\em
half-parabolic} potential.
\item $\alpha \beta  = \zeta \neq 0$ and  $\beta \rightarrow 0$: {\em
    Urabe's potential}.
\begin{equation} \label{Urabpot}
V(x)   =    \frac{\omega^2}{2\zeta^2}   \left(   1   -\sqrt{1+2\zeta
  x}\right)^2 \ \ \ x \in [-\frac{1}{2\zeta},\frac{3}{2\zeta}]
\end{equation}
Notice that  in this last case,  the condition $\alpha  \in [0,1]$ has
been  relaxed. Preserving  $|S(X)|  < 1$  amounts  to restricting  the
values of  $X$ to $|X| <  1/|\zeta|$. Hence, a potential defined only 
on a finite interval of the real axis.
\end{itemize}

Note that  the first and third  results can be obtained  by means of
the general relations (\ref{Vbeta0}) and (\ref{Vbetainf}).

\subsubsection{Family II} \label{ssIIC2}

Let
\begin{equation} \label{SDor}
S(X) =  \frac{1}{\xi}\frac{\sinh X}{\cosh X  - 1 +  \alpha}\, , \  \ \
(\alpha,\xi) \in \R^{*+} \, .
\end{equation}
$S(X)$  is  readily odd.  According  to  the  value of  $\alpha$,  the
condition $|S(X)|<1$ imposes
\begin{equation} \label{xiofalpha} 
\left\{  \begin{array}{ccc} 0  <  \alpha <  1  & \Rightarrow  & \xi  >
[\alpha (2-\alpha)]^{-1/2} \\ \alpha \geq 1 & \Rightarrow & \xi \geq 1
\end{array} \right.
\end{equation}
The  shear  function of  (\ref{SDor})  is  given by  $\bar{x}(X)=\{\ln
[(\cosh(X)-1+\alpha)/\alpha]\}/\xi$   and   inverting   (\ref{relxXS})
(or \eqref{xofX}) amounts to solving
\begin{equation} \label{Yeq}
Y^{\xi+1} + 2(\alpha-1)Y^{\xi} + Y^{\xi-1} -2\alpha e^{\xi x} = 0
\end{equation}
where $Y = \exp X$.  For some particular values of $\alpha$ and $\xi$,
the above algebraic equation can be solved exactly for $Y$. This leads
to an  analytical expression  for the potential.  In what  follows, we
work  out some  of the  simplest  cases and  systematically apply  the
scaling (\ref{scaledVsamefreq}) to derive the corresponding family. \\

\noindent
$\bullet$ $\xi = 1$ and $\alpha \geq 1$.
\begin{eqnarray} \label{Vxi1}
V(x)   &=&  \frac{\omega^2}{2\beta^2}   \left[  \ln   \left(  1-\alpha
    +\sqrt{2\alpha    (e^{\beta   x}-1)+\alpha^2}\right)   \right]^2\,
    ,\nonumber \\ & & x > - \frac{1}{\beta}\ln 2\alpha\, .
\end{eqnarray}
This represents a two-parameter family of potentials singular in $x =
- (\ln 2\alpha)/\beta $ and defined on a half-line.\\ Notice that for
$\alpha=1$, the  analytical form of this  potential simplifies further
to yield the remarkably simple expression
\begin{equation} \label{Vxi1a1}
V(x) = \frac{\omega^2}{8\beta^2}  \ln^2 \left( 2e^{\beta x}-1\right)\,
    ,\ x > - \frac{\ln 2}{\beta}\, .
\end{equation}

\noindent
$\bullet$ $\xi = 2$ and $\alpha = 1$.
\begin{eqnarray} \label{Vxi2alpha1}
V(x)  &=&  \frac{\omega^2}{2\beta^2}  \left[\frac{2}{3}\beta x  +  \ln
    \left(   q_+^{1/3}  -   q_-^{1/3}\right)\right]^2\, ,\\   q_{\pm}  &=&
    \left(\sqrt{1+\frac{e^{-4\beta x}}{27}} \pm 1 \right) \, .\nonumber
\end{eqnarray}
Contrary to the previous family, this one is defined on
the whole line $\R$ and is not singular. \\

\noindent
$\bullet$ $\xi = 3$ and $\alpha = 1$.
\begin{eqnarray} \label{Vxi3alpha1}
V(x)      &=&      \frac{\omega^2}{8\beta^2}     \left[\ln      \left(
  \frac{\sqrt{1+8e^{3\beta x}}-1}{2} \right)\right]^2\, ,
\end{eqnarray}
is defined on the whole line $\R$  and is not singular, as in the previous
case. \\

Due to Cardano's formula, \eqref{Yeq} can be solved 
analytically in the more general case where $\xi$ is either equal  
to $2$ or $3$ and $\alpha  > 1$. For the same reason, it is also possible 
to derive exact solutions of \eqref{Yeq} when
$\alpha = 1$ and $\xi=5$ or $7$.  
These expressions, however, are somewhat messy and are not provided.
It is also possible to derive exact solutions of \eqref{Yeq} 
when $\alpha=2$ and $\xi=5$ or $7$. The potentials given above are  
a few among many others.

Whatever the values of the parameters $\xi$ and $\alpha$, the limiting
case $\beta \rightarrow  0$ leads to the harmonic  potential as $S(X)$
given by  (\ref{SDor}) is continuous in $X=0$.  Given that 
$\langle S \rangle =  1/\xi$, the limiting
case $\beta  \rightarrow \infty$ gives rise to a split-harmonic oscillator 
whose left and  right frequencies are
$\omega_{\pm} =  \xi \omega  /(\xi \pm 1)$.  In the first  case, where
$\xi = 1$, this leads to a half-parabolic well.
 
\section{Beyond EBK} \label{WKB2nd}
We  now  turn to  the  problem of  the  determination  of the  quantum
spectrum  of   isochronous  potentials.  As  already   stated  in  the
introduction, their semiclassical EBK spectrum (i.e., obtained from
a first order WKB method) is
perfectly  regularly  spaced.  Nevertheless,  a  study  of  the  exact
spectrum of the split-harmonic oscillator leads to the conclusion that
at least  not all isochronous potentials  possess strictly equidistant
energy  levels (\cite{Still89,Ghosh81}).  As the  determination  of an
exact  quantisation condition is  not possible  in general,  a natural
idea is  to go  beyond the usual  semiclassical approximation  and to
study  the properties of  the spectrum  by means  of the  higher order
terms generated by the WKB method.

The  WKB method  to  all orders  has  been first  developed by  Dunham
\cite{Dun32}    and   subsequently    improved    by   many    authors
\cite{Bend77,Robnik97,Roman00,Robnik00}.  In  particular, it allows  
the quantisation  condition for  a 1D  analytic potential  to be written 
as a power  series  in $\hbar$.  Such  series  have  to be  interpreted  as
asymptotic series and are  generally not convergent. However, 
the result leads to  the exact
quantisation condition for those exactly solvable potentials, 
whose  WKB series  can be evaluated explicitly  and summed.

In what follows,  we briefly recall the WKB method  before to make use
of the second and fourth terms derived in \cite{Robnik97} to express
the  first corrections  to the  EBK quantisation  
in terms of  the $S$ function presented  in the previous section.

\subsection{WKB to all orders}

For a  complete description  of the method  and the properties  of WKB
series,         the        reader        is         referred        to
\cite{Bend77,Robnik97,Roman00,Robnik00}.  The present  introduction is
mainly inspired  by \cite{Robnik97}.  We start from  the Schr\"odinger
equation in 1D ($m=1$),
\begin{equation} \label{Schrod}
\left[ -\frac{\hbar^2}{2} \frac{d^2}{dx^2} +  V(x)\right] \psi (x) = E
\psi (x)\, .
\end{equation}
Writing
\begin{equation} \label{Psi}
\psi (x) = \exp \left( \frac{i}{\hbar} \sigma (x)\right)\, ,
\end{equation}
we obtain
\begin{equation} \label{diffsig}
\sigma'^{\, 2} (x)  + \left( \frac{i}{\hbar} \right) \sigma''  (x) = 2
(E-V(x))
\end{equation}      
solved by means of a power series in $\hbar$,
\begin{equation} \label{sersig}
\sigma  (x)  =  \sum_{k=0}^{\infty} \left(  \frac{\hbar}{i}  \right)^k
\sigma_k (x) \, .
\end{equation} 
Reinstating  (\ref{sersig}) in  (\ref{diffsig}) and  solving  order by
order in $\hbar$ yields the recurrence
\begin{equation} \label{sigrec}
\sigma_0'^{2}  = 2(E-V) \  \ \text{and}  \ \  \sum_{k=0}^{l} \sigma'_k
\sigma'_{l-k} + \sigma'_{l-1} = 0 \, , \ l \geq 1
\end{equation} 
Requiring the  wave function to  be single valued amounts  to imposing
the following quantisation condition
\begin{equation} \label{Quantcond}
 \oint_{\gamma}   \!    d   \sigma   =    \sum_{k=0}^{\infty}   \left(
 \frac{\hbar}{i} \right)^k  \oint_{\gamma} \! d \sigma_k  = 2\pi \hbar
 n\, ,\ n \in \N
\end{equation}
where  the  complex integration  contour  $\gamma$  surrounds the  two
turning points of $V(x)$ at energy $E$ located on the real axis.

The first term of the series (\ref{Quantcond}) is readily proportional
to the classical action $I_0(E)$
\begin{equation} \label{intsigma0}
 \oint_{\gamma}   \!   d   \sigma_0   =   2   \int_{x_-(E)}^{x_+(E)}\!
 \sqrt{2(E-V(x))} \, dx = 2\pi I_0(E) \, .
\end{equation}
As integral of a logarithmic  derivative, the second term can be shown
to be equal to
\begin{equation} \label{intsigma1}
\left(  \frac{\hbar}{i} \right)  \oint_{\gamma} \!  d \sigma_1  = -\pi
\hbar \, .
\end{equation}
If we  truncate the WKB  series to this  last order, we  re-derive the
so-called EBK quantisation
\begin{equation} \label{EBK}
I_0 (E) = \left( n+\frac{1}{2}\right) \, \hbar \, ,\ n \in \N \, ,
\end{equation}
where the $1/2$ term represents the Maslov index.

As shown by Fr\"oman  \cite{Froman}, all odd terms $\sigma'_{2k+1}$, $k
\geq  1$,  are  total  derivatives  and as  such,  their  contribution
vanishes.   This  allows us  to  rewrite   the   quantisation  condition
(\ref{Quantcond}) as
\begin{equation} \label{Quantcond2}
 \sum_{k=0}^{\infty} I_{2k} (E) = \left( n+\frac{1}{2}\right) \, \hbar
 \, ,\ n \in \N
\end{equation}
where we have defined
 \begin{equation} \label{I2k}
 I_{2k}  (E)  =  \frac{1}{2\pi}  \left(  \frac{\hbar}{i}  \right)^{2k}
\oint_{\gamma} \! d \sigma_{2k}\, , \ k \in \N \, .
\end{equation}    

In  case $V(x)$ is  analytic and  $V'(x) \neq  0$ if  $x \neq  0$, the
authors  of \cite{Robnik00}  have  proved that  the contour  integrals
involved in  (\ref{I2k}) can be systematically  replaced by equivalent
Riemann  integrals between  the  two turning  points.  Thanks to  this
result it can be shown that \cite{Robnik97}

\begin{equation} \label{I2}
 I_2 (E)  = -\frac{\hbar^2}{24\sqrt{2}\pi} \frac{\partial^2 }{\partial
 E^2}         \int\limits_{x_-(E)}^{x_+(E)}         \!\!\!        dx\,
 \frac{V'^2(x)}{\sqrt{E-V(x)}}
\end{equation}
and

\begin{eqnarray} \label{I4}
 I_4   (E)  &=&   \frac{\hbar^4}{4\sqrt{2}\pi}   \left[  \frac{1}{120}
\frac{\partial^3 }{\partial  E^3} \int\limits_{x_-(E)}^{x_+(E)} \!\!\!
dx\,   \frac{V''^2(x)}{\sqrt{E-V(x)}}  \right.   \nonumber  \\   &  -&
\left.     \frac{1}{288}      \frac{\partial^4     }{\partial     E^4}
\int\limits_{x_-(E)}^{x_+(E)}      \!\!\!      dx\,      \frac{V'^2(x)
V''(x)}{\sqrt{E-V(x)}} \right] \, .
\end{eqnarray}

As the $I_{2k}$'s are  originally given by contour integrals, addition
of  total  derivatives  to  the $\sigma'_{2k}$  doesn't  change  their
value. However, it modifies their expression in terms of the potential
and its derivatives and possibly allows for their simplification. Such
a    technique    is   developed    and    used   systematically    in
\cite{Robnik00}.  Expressions (\ref{I2}) and  (\ref{I4}) are one
among others (see for instance \cite{Vran00}).

\subsection{WKB expansion for isochronous potentials}

We  now use  the results  obtained in  section \ref{Gener}  to express
$I_2$  and  $I_4$ in  terms  of the  function  $S(X)$  related to  the
potential $V(x)$. We make use of the change of variables $V(x)=\omega^2
X^2/2$ and of the relation (\ref{reldxdXS}) to find
\begin{equation} \label{VprimeX}
\frac{d V}{d x} = \frac{\omega^2 X}{1 + S(X)} \, .
\end{equation}
Remarking further that $X(x_{\pm} (E)) = \sqrt{2E}/\omega$, we obtain
\begin{equation} \label{I2X}
 I_2 (E)  = -\frac{\hbar^2 \omega^4}{24\sqrt{2}  \pi} \frac{\partial^2
 }{\partial                          E^2}                         \!\!
 \int\limits_{-\frac{\sqrt{2E}}{\omega}}^{\frac{\sqrt{2E}}{\omega}} \!
 \frac{X^2 dX}{(1+S(X)) \sqrt{E-\frac{1}{2}\omega^2 X^2}} \, .
\end{equation}
Then using the fact that $S(X)$  is odd and making the change of variables 
$u=\omega^2 X^2/2$, we find 
\begin{equation} \label{I2u}
 I_2 (E) =  -\frac{\hbar^2 \omega}{12 \pi} \frac{\partial^2 }{\partial
 E^2} \left[  E \int\limits_{0}^{1} \!  \, \frac{u^{1/2}}{(1-u)^{1/2}}
 \frac{du}{1-S^2\left(\frac{\sqrt{2Eu}}{\omega}\right)} \right] \, .
\end{equation}

Another equivalent formulation suitable for an asymptotic analysis
is given by
\begin{equation} \label{I2v}
 I_2    (E)   =    -\frac{\hbar^2   \omega}{12    \pi}   \frac{1}{E^2}
  \int\limits_{0}^{E} \!  \, \frac{v^{3/2}}{(E-v)^{1/2}} \frac{d^2 }{d
  v^2}                                                           \left[
  \frac{v}{1-S^2\left(\frac{\sqrt{2v}}{\omega}\right)}\right] dv \, .
\end{equation}
This   way,   $I_2$   is   expressed   through  an   Abel type   (or
Riemann-Liouville  fractional) integral  about  which a  lot is  known
\cite{Bleistein,Erdel}.

A similar calculation yields the fourth order correction
\begin{eqnarray} \label{I4u}
 I_4   (E)   &=&   \frac{\hbar^4}{4\pi\omega}   \left[   \frac{1}{120}
\frac{\partial^3     }{\partial     E^3}    \int\limits_{0}^{1}     \!
\frac{G_1\left(\frac{\sqrt{2Eu}}{\omega}\right)
}{u^{1/2}(1-u)^{1/2}}\,    du    \right.    \nonumber    \\    &    -&
\left.     \frac{1}{288}      \frac{\partial^4     }{\partial     E^4}
\int\limits_{0}^{1}                                                  \!
\frac{G_2\left(\frac{\sqrt{2Eu}}{\omega}\right)
}{u^{1/2}(1-u)^{1/2}}\, du \right] \, .
\end{eqnarray}
where the functions $G_1$ and $G_2$ are given by
\begin{eqnarray} \label{G12}
G_1(X)     &=&    \frac{\omega^4}{(1-S^2)^3}    \left[     3S^2+1    +
  \frac{8XSS'(1+S^2)}{1-S^2}     \right.      \nonumber     \\     &+&
  \left.   \frac{X^2S'^2(1+10S^2+5S^4)}{(1-S^2)^2}    \right]   \,   ,
  \nonumber \\ G_2(X)  &=& \frac{\omega^6X^2}{(1-S^2)^3} \left[ 3S^2+1
  + \frac{4XSS'(1+S^2)}{1-S^2} \right]\, . 
\end{eqnarray}
In the last expressions, $S$ stands for $S(X)$ and $S'$ for $dS/dX$.

Similar to  $I_2$, $I_4$ is equivalently expressed  through Abel 
integrals via
\begin{eqnarray} \label{I4v}
 I_4   (E)   &=&   \frac{\hbar^4}{4\pi\omega}   \left[   \frac{E^{-3}}{120}
\int\limits_{0}^{E}   \!   \frac{v^{5/2}}{(E-v)^{1/2}}
\frac{d^3}{dv^3}\left\{G_1\left(\frac{\sqrt{2v}}{\omega}\right)\right\}
dv \right.   \nonumber   \\  &  -&   \left.  \frac{E^{-4}}{288}
\int\limits_{0}^{E}     \!    \frac{v^{7/2}}{(E-v)^{1/2}} \frac{d^4}{d v^4}
\left\{G_2\left(\frac{\sqrt{2v}}{\omega}\right)\right\} dv \right] \, .
\end{eqnarray}
  
\subsection{Scaling properties of WKB series}

We  have  seen in  section  \ref{scalesec}  that  the general  scaling
$\tilde{V}(x)=(\gamma/\beta)^2   V(\beta   x)$,  $(\gamma,\beta)   \in
\R^{*2}$, preserves  isochronism and changes the  frequency $\omega$ of
$V(x)$  to $\tilde{\omega}=\gamma  \omega$.  Such  a transformation,
however, is  not expected to  have simple consequences at  the quantum
level given  the lack  of relation between  the spectra of  $V(x)$ and
$\tilde{V}(x)$ in  the general case.   This can be seen  directly from
the Schr\"odinger equation itself
\begin{eqnarray} 
\left[ -\frac{\hbar^2}{2}  \frac{d^2}{dx^2} + \tilde{V}(x)\right] \psi
(x)  &=&   \tilde{E}  \psi  (x)  \label{SchrodV}   \\  
\Leftrightarrow  \left[
-\frac{\hbar^2}{2}                  \frac{d^2}{dy^2}                 +
\frac{\gamma^2}{\beta^4}V(y)\right]  \varphi  (y)  &=&  
\frac{\tilde{E}}{\beta^2} \varphi (y) \label{SchrodVtilde}
\end{eqnarray}
where $y=\beta  x$ and  $\varphi(y) = \psi(x)$.\\  This last  equation is
equivalent to  the Schr\"odinger  equation for $V$  iff $\gamma  = \pm
\beta^2$, which  corresponds to the particular  scaling $\tilde{V}(x) =
\beta^2 V(\beta x)$. The spectrum  of $\tilde{V}$ is then the spectrum
of $V$ multiplied by $\beta^2$.

But, nothing can be said about the general transformation and  the
frequency-preserving scaling  $\tilde{V}(x) = V(\beta  x )/\beta^2$ in
particular without further information regarding the potential $V(x)$.

As we  are dealing with  isochronous potentials,  we may
transform the Schr\"odinger equation (\ref{Schrod}) into an equivalent
Sturm-Liouville problem involving the  function $S(X)$ rather than the
potential $V(x)$ itself. Thanks to (\ref{Xdef}) and (\ref{reldxdXS}), 
we obtain
\begin{equation} \label{SturmLiouv}
\frac{\hbar^2}{2}\frac{d}{dX} \left(\frac{1}{1+S}\frac{d\phi}{dX}\right)+
 (1+S)\left[E-\frac{\omega^2}{2}X^2 \right] \phi = 0
\end{equation}
where $\phi  \equiv \phi(X)=\psi(x)$ and  $S \equiv S(X)$. This last
form proves convenient for numerical purposes as $S(X)$ is a
supplied function whose analytical form can be chosen freely within
the requirements (\ref{Sodd}) and (\ref{Smod}). But we found it
of no particular help in investigating scaling effects on the spectrum.
 
Instead, we can look at the way this scaling affects each term of the WKB
series. 
\begin{claim}
Let $I_{2n}$ and $\tilde{I}_{2n}$ be the  $2n^{\rm th}$ terms of the
WKB series for the potentials $V(x)$ and 
$\tilde{V}(x) = (\gamma/\beta)^2 V(\beta
x)$ respectively. Then,
\begin{eqnarray} \label{WKBscaling}
\tilde{I}_{2n}   (E)  =   \left(  \frac{\beta^2}{\gamma}\right)^{2n-1}
I_{2n} \left( \frac{\beta^2 E}{\gamma^2}\right)\, ,\ n \in \N .
\end{eqnarray}
\end{claim}
Notice that this expression is {\em valid for any potential} and not
only for isochronous potentials. Details of the proof are given
in appendix \ref{App1}. But let us immediately remark that, multiplying 
equation \eqref{SchrodVtilde} by $\beta^4/\gamma^2$ indicates that the 
energy is rescaled by $\beta^2/\gamma^2$ and $\hbar$, by 
$\beta^2/\gamma$. Since each term of the WKB series is proportional to a 
given power of $\hbar$, the above result is easily established.

Thus, we arrive to the conclusion that, even though no simple transform
enables us to derive the  spectrum of $\tilde{V}$ from the spectrum of
$V$, each  term of the WKB  series obeys its own  scaling, which allows
for calculating  the WKB series for  $\tilde{V}$ once it  is known for
$V$. We may restrict our 
investigation to potentials with a fixed frequency and a fixed value 
for $\beta$.

\section{Application} \label{Twoparafamily}

We  now  apply  what  precedes  to  some  isochronous
potentials presented in section  \ref{Iso}. We start our investigation
with  the   family  of  potentials  derived  by   Bolotin  and  MacKay
\cite{Bol02} and  obtain an analytical expression for its fourth order 
WKB quantisation condition. This result is immediately generalised to
the two-parameter Stillinger's family (family $I$) thanks to the scaling
(\ref{WKBscaling}).

Regarding the  class of potentials presented  in \ref{ssIIC2} (family $II$), 
we  won't be able to derive any explicit analytical expression
for  the  corrections. We  will nevertheless derive  their  asymptotic
behaviour.

\subsection{Family $I$}

For the sake of simplicity, we calculate the terms $I_2$ and $I_4$ for
a potential  of frequency  $\omega = \sqrt{2}$  whose $S$  function is
given by
\begin{equation}
S(X) = \sqrt{1-\eta} \frac{X}{\sqrt{1+X^2}}\, ,\ \ \eta \in [0,1].
\end{equation}
We recover  a general  expression of the  type (\ref{VStill})  for the
potential by means  of (\ref{scaledxbarS}) with $\tilde{S}(X)=S( \beta
\sqrt{\alpha} X)$ and $\eta=1-\alpha$.  The frequency is then fixed to
the value $\omega$ thanks to $\gamma = \omega/\sqrt{2}$.

\subsubsection{Analytic expressions for $I_2$ and $I_4$}

Setting $\hbar =1$, expression (\ref{I2u}) leads to
\begin{equation} \label{I2bol} 
I_2 (E) = - \frac{\sqrt{2}}{2^4} \frac{(1-\eta)}{(1+\eta E)^{5/2}}
\end{equation}  
and expression (\ref{I4u}) to
\begin{equation} \label{I4bol} 
I_4  (E) =  \frac{\sqrt{2}}{2^{10}}  \frac{(1-\eta) p(\eta,E)}{(1+\eta
  E)^{15/2}}
\end{equation} 
where
\begin{eqnarray}
 p(\eta,E) &=&  15 \eta^4 E^4  -30 \eta^3 (7\eta-16) E^3  \nonumber \\
&-&  3   \eta^2  (119   \eta^2-651\eta+411)  E^2\nonumber  \\   &+&  3
\eta(455\eta^2-567\eta+104) E -280 \eta^2 +140 \eta + 8 \nonumber
\end{eqnarray}

\subsubsection{ WKB quantisation condition to fourth order} \label{sssWKBQCFI}

The fourth order WKB quantisation condition  (\ref{Quantcond2}) for
the family of potentials  (\ref{VStill}) is deduced from (\ref{I2bol})
and (\ref{I4bol}) and from the scaling relation (\ref{WKBscaling})
\begin{equation} \label{Quantbol}
E         -\frac{1}{8}\frac{(\alpha        \beta)^2}{Q^{5/2}}        +
\frac{\alpha^4\beta^6}{2^8\,  \omega^2} \frac{P}{Q^{15/2}}  =\left(n +
\frac{1}{2}\right) \omega \, .
\end{equation}
where
\begin{equation} \label{PandQ}
Q = \left[1+2\alpha  (1-\alpha) \frac{\beta^2}{\omega^2} E \right]\ ;\
P     =     p\left(1-\alpha,2\alpha     \frac{\beta^2}{\omega^2}     E
\right)
\end{equation}  

We  can  now solve  equation  (\ref{Quantbol})  for  $E$ to  obtain  a
semiclassical approximation of the spectrum. This approximation should
be  valid  asymptotically,  that is,  as  $E  \rightarrow
\infty$, as we  expect $I_0(E) \ll I_2(E) \ll I_4(E)  \cdots $. 
We can check  that this is indeed the case for  $\alpha < 1$ as
$I_0(E)  \propto E$,  $I_2(E)  \propto E^{-5/2}$  and $I_4(E)  \propto
E^{-7/2}$.  Notice that  for  the particular  value  $\alpha =0$  (or
$\beta=0$), the potential  (\ref{VStill}) becomes a harmonic potential
and the corrections  $I_2$ and $I_4$ vanish. It can  be shown that all
higher  corrections  $I_{2n}$   vanish  as  well  \cite{Bend77}, which
establishes the well known fact that, in this case, 
the EBK quantisation leads to an exact result.

However, for  $\alpha  =1$,  the  potential  (\ref{VStill}) becomes  a
singular isotonic potential (see  section \ref{ssIIC1}). The
second  and  fourth  corrections  become  energy-independent  in  this
limit. It can  be shown by calculating the entire  WKB series that all
corrections are energy-independent and  that their summation leads to
the  exact  quantisation  condition \cite{Roman00,Barclay94}.  We  can
check that the  two corrective terms $I_2$ and  $I_4$ we have obtained
are  the coefficients  of the  Taylor expansion  for  the non-integral
Maslov  index  (see  \cite{Friedrich96})  of  the  exact  quantisation
condition in $\beta = 0$. Indeed, the exact quantisation condition for
the   radial  harmonic   oscillator   (\ref{Isotonic})  (or   isotonic
potential) reads (see for example \cite{Nieto79})
\begin{equation}
E_n = (n+\frac{\mu}{4})\, \omega \nonumber
\end{equation}
where the non-integral Maslov index is given by
\begin{equation} \label{Maslovisotonic}
\mu           =          3           +          \frac{\omega}{\beta^2}
\left(\sqrt{1+\frac{\beta^4}{\omega^2}}-\frac{\beta^2}{\omega}-1
\right).
\end{equation}
Expanding this last expression around $\beta=0$ yields
\begin{equation}
E_n       \sim       (n+\frac{1}{2}+\frac{1}{8}\frac{\beta^2}{\omega}-
\frac{1}{32}\frac{\beta^6}{\omega^3} + \cdots)\, \omega \, ,\nonumber
\end{equation}
which is  precisely what we  obtain from (\ref{Quantbol}) given that,
$\alpha=1$ implies $Q=1$ and $P=8$.

This remark brings  us to consider that WKB series  should be more and
more accurate as the scaling  parameter $\beta$ tends to zero, i.e. in
the limit where  the potential tends to the  harmonic oscillator. This
statement is confirmed by  the numerical computations performed in the
next sections.

\subsection{Family II} \label{FamilyII}

We restrict our study to the nonsingular potentials \eqref{Vxi2alpha1} 
and \eqref{Vxi3alpha1} for which the WKB method leads to an asymptotic
series as $E \rightarrow \infty$ 
\footnote{For the family of singular potentials \eqref{Vxi1a1}, the 
corrections $I_2(E)$ and $I_4(E)$ can be calculated analytically. 
For example, setting $\hbar=1$, 
$\beta=1$ and $\omega=\sqrt{2}$ in \eqref{Vxi1a1}, we obtain 
$I_2(E) = -\frac{\sqrt{2}}{48}
[{\sf I}_0(2\sqrt{E})+\frac{{\sf I}_1(2\sqrt{E})}{2\sqrt{E}}]$,
where ${\sf I}_n(z)$
is the modified Bessel function of order $n$ as defined in 
\cite{Abramowitz}. A similar expression involving the ${\sf I}_n$'s 
up to order 3 can be obtained for $I_4(E)$. As $z \rightarrow \infty$, 
${\sf I}_n(z)\sim e^z/\sqrt{2\pi z}$ (see formula 9.7.1 in \cite{Abramowitz}).
Therefore, $I_2(E)$ and $I_4(E)$ grow exponentially fast as 
$E \rightarrow \infty$. We can see that they alternate in sign though. 
It is not clear to us whether the entire WKB series could be summed 
and would eventually be finite as $E \rightarrow \infty$. Numerical
results for \eqref{Vxi1a1} seem to suggest that 
$E_n \sim (n+\frac{\mu(\beta)}{4})\hbar \omega$, when 
$n \rightarrow \infty$, with a Maslov index, $\mu(\beta)$, 
ranging from $2$ (small $\beta$) to $3$ (large $\beta$).}.  
In what follows, we again evaluate $I_2$  and $I_4$  for $\omega =
\sqrt{2}$.  The $S$ function of \eqref{Vxi2alpha1} and \eqref{Vxi3alpha1}
reads
\begin{equation} \label{SVxialpha1}
S_{\xi}(X) = \frac{1}{\xi} \tanh(X)\, ,\ \ \xi \in \{2,3\}.
\end{equation}
We recover the  general expression for the potentials  by means of the
scalings \eqref{genscaling} and \eqref{scaledxbarS}.  The frequency is
brought back to the general value $\omega$ by choosing 
$\gamma = \omega/\sqrt{2}$.

\subsubsection{Expressions for $I_2$ and $I_4$}

Although $S_{\xi}(X)$ in \eqref{SVxialpha1} has a very simple form, we
are  not able  to provide  an analytical  expression for  $I_2(E)$ and
$I_4(E)$.
Then, we resort to their numerical  evaluation as a function of $E$ and
use  the  corresponding   functions  to  calculate  the  semiclassical
spectrum.

The lack of exact analytical  forms for $I_2(E)$ and $I_4(E)$ does not
prevent  us from  extracting some  useful information  regarding their
asymptotic behaviour as $E \rightarrow \infty$.  According to the fact
that WKB should provide the best results in this limit, we expect this
analysis  to yield  the right  correction to the harmonic levels  at high
energy. Using  \eqref{I2v} and \eqref{I4v} and a  result regarding the
asymptotic   expansion    of   fractional   integrals    obtained   in
\cite{Berger75} (see also 
\footnote{The result contained in \cite{Berger75} is actually not correct.
For $f \sim e^{-\alpha t}\sum_{m=0}^{\infty} d_mt^{-r_m},\ t \rightarrow 
\infty$, and $\alpha>0$, the generalised fractional integral
$$I^{\mu}_{\lambda^p}f (\lambda) = 
\frac{1}{\Gamma(\mu)}\int_0^{\lambda}(\lambda^p-\xi^p)^{\mu-1}
p\xi^{p-1}f(\xi)d\xi $$ has an asymptotic behaviour given by
$$I^{\mu}_{\lambda^p}f (\lambda) \sim \sum_{n=0}^{\infty} 
\frac{p M[f;p(n+1)]}{n!\Gamma(\mu-n)} (-1)^n\lambda^{-p(n-\mu+1)} $$
as $\lambda \rightarrow \infty$. 
Here, $M[f;x]=\int_0^{\infty}\!\!f(u)u^{x-1}du$ 
is the Mellin transform of $f$. 
In \cite{Berger75}, the first factor $p$
appearing in the series above is missing due to an error in the calculation
of some residue. The reader is referred to this paper for more details. 
This correct asymptotic expansion has been used to calculate the 
coefficients of table \ref{table1}.}), 
we easily derive the large-energy behaviour of $I_2(E)$ and $I_4(E)$ 
to be

\begin{eqnarray} \label{asymptI2xi2}
I_2(\xi;E)                                                       &\sim&
\frac{M_{2,1}(\xi)}{E^{5/2}}+\frac{M_{2,2}(\xi)}{E^{7/2}}+
o\left(\frac{1}{E^{7/2}}\right)   \nonumber   \\   I_4(\xi;E)   &\sim&
\frac{M_{4,1}(\xi)}{E^{7/2}}+ o\left(\frac{1}{E^{7/2}}\right)
\end{eqnarray}
where  $\xi   \in  \{2,3\}$   and  $M_{m,n}(\xi)$  are   some  nonzero
coefficients  related  to  the   Mellin  transform  of  the  functions
appearing in the numerator  of the  fractional integrals  involved in
\eqref{I2v} and \eqref{I4v},  (see \cite{Bleistein} or \cite{Berger75}
for more  details). The numerical  values of these coefficients  for a
frequency $\omega=\sqrt{2}$ are listed in table \ref{table1}.

\begin{table}
\caption{Table  of   the  coefficients  $M_{m,n}(\xi)$   appearing  in
  \eqref{asymptI2xi2} for $\omega=\sqrt{2}$.\label{table1}}
\begin{ruledtabular}
\begin{tabular}{ccc}
$M_{m,n}(\xi)$  & $\xi=2$  & $\xi=3$  \\ \hline  $M_{2,1}$  & 
$1.9020\times 10^{-2}$ & $6.3076\times 10^{-3}$ \\ $M_{2,2}$ &  
$1.7287\times 10^{-1}$ &
$5.5608\times 10^{-2}$ \\ $M_{4,1}$ &  $7.0128\times 10^{-3}$  & $1.9414\times
10^{-3}$
\end{tabular}
\end{ruledtabular}
\end{table}

\subsubsection{Asymptotic quantisation condition}

Our asymptotic expansion  in  $E$ is truncated at  order
$E^{-7/2}$ in  \eqref{asymptI2xi2} because the term  $I_6(E)$ would be
needed to evaluate the proper  coefficient of order $E^{-9/2}$ for the
entire  WKB series.  We note again on this example that WKB series are
generally no 
{\em direct} asymptotic expansions in energy. Obtaining such 
a series requires $I_{2n}(E)$ to be expanded as
asymptotic series before terms of equal power in $E$ can be collected.
A method has  been developed  in \cite{nanayakkara01}  which  allows 
for  the  derivation  of  a proper  (direct) asymptotic
expansion in $E$.  Its range of application seems  to be restricted to
power-law potentials, however, and  it is not  clear how it  could be
used in the present case (see \cite{nanayakkara02,nanayakkara03}).

Finally,  applying the scaling \eqref{WKBscaling}  to \eqref{asymptI2xi2},
we  obtain   an  asymptotic  quantisation   condition  for  potentials
\eqref{Vxi2alpha1}  and  \eqref{Vxi3alpha1}  valid for  any  frequency
$\omega$ and  any parameter $\beta$ such that  $(\beta/\omega)^2 E$ is
large
\begin{multline} \label{asymptQCxi23}
E+\frac{\omega^5}{4\beta^3}\frac{M_{2,1}}{E^{5/2}}+
\frac{\omega^5}{4\beta}
\left\{\frac{\omega^2}{2\beta^4}M_{2,2}+M_{4,1}\right\}\frac{1}{E^{7/2}}
\\  \sim  (n+\frac{1}{2})\omega \,  ,\  \ \  \frac{\beta^2E}{\omega^2}
\rightarrow \infty .
\end{multline}
For simplicity, we have dropped  the $\xi$ dependence of
the    coefficients   $M_{m,n}(\xi)$.     The   last    condition   in
\eqref{asymptQCxi23} is  absolutely essential. It comes  from the fact
that, when deriving the  asymptotic forms \eqref{asymptI2xi2}, we have
assumed  that  $E$  was  large   in  $I_2(E)$  and  $I_4(E)$.  As  
\eqref{WKBscaling} rescales the energy by a factor $(\beta/\omega)^2$,
we  have to  consider  that $(\beta/\omega)^2  E$  is now large.  For
example,  this condition ensures  that the  corrective terms  of order
$E^{-5/2}$  and $E^{-7/2}$  tend  to zero  in \eqref{asymptQCxi23}  as
$\beta \rightarrow 0$ (harmonic limit) providing $E > \beta^{-2}$.

\section{Numerical evaluation of the spectrum}

\subsection{Numerical solution of singular Sturm-Liouville problems}

As  the  difference between  the  spectrum  of  a general  isochronous
potential and the harmonic one  is expected to be small, its numerical
determination requires an accurate method.

Our   numerical  evaluation   of  the   spectrum  is   based   on  the
Sturm-Liouville  problem -  SLP -  (\ref{SturmLiouv}). We  have solved
this  equation by  means of  a shooting  method provided  by  the code
SLEIGN2 developed by  Bailey, Everitt and Zettl \cite{Bailey01,Zettl}.
Among other  advantages, solving (\ref{SturmLiouv})  only requires the
analytical expression  for $S(X)$ and  not the  potential. It
can then be easily used for a wide range of potentials.

The SLP (\ref{SturmLiouv}) is always singular given that its endpoints
are  located  at  $X  =  \pm \infty$  (where  the  function  $\phi(X)$
vanishes).   The  type  of  singularity  may  vary  according  to  the
potential under consideration. But in any case, it is the same for the
SLP  (\ref{SturmLiouv})  and  the  associated  Schr\"odinger  equation
(\ref{SchrodV}) (see \cite{Fulton} lemma 1 and 2).

\subsection{Endpoint classification} 

Runing  SLEIGN2 requires  knowledge of the singularity type of each
endpoint of the SLP under investigation. We,then, classify 
hereafter those susceptible to be met  in handling isochronous potentials.  
To  determine the type
of singularity of each branch of $V(x)$, we first remark that
\begin{claim} \label{cineqiso}
If $V(x)$ is isochronous of frequency $\omega$
\begin{equation} \label{ineqiso}
\forall x \in \R,\ \ \ V(x) \geq \frac{\omega^2}{8} x^2
\end{equation}
\end{claim}

This  is a  consequence of  the fact  that $|S(X)|  < 1  \ \Rightarrow
|\bar{x}(X)| \leq |X|$. Now, $x = X + \bar{x}(X)\ \Rightarrow |x| \leq
2|X|$ hence $\omega^2 x^2 /8 \leq \omega^2 X^2 /2 = V(x)$.

\begin{claim}
An endpoint $x_e$  of $V(x)$ located at $\pm \infty$  is always of both 
the {\em limit-point} and {\em nonoscillatory} type.
\end{claim}
Indeed, applying  theorem 7 of \cite{Fulton}  and claim \ref{cineqiso}
proves this endpoint to be of the {\em limit-point} type. Moreover, as
$\lim_{x \rightarrow  x_e} V(x) =  \infty$, applying theorem 6  of the
same paper shows that $x_e$ is of the {\em nonoscillatory} type.

\begin{claim}
Assume an endpoint $x_e$ of $V(x)$ is located at $x_0$, where $x_0$ is
finite.   Then it  is always  of the  {\em nonoscillatory}  type.\\ 
$\bullet$ If
$\displaystyle \lim_{x \rightarrow x_0}  \inf \, (x-x_0)^2 V(x) > 3/8$
or\\  if   $(x-x_0)^2  V(x)  \geq  3/8$   and  $\displaystyle  \lim_{x
\rightarrow x_0} \inf \, (x-x_0)^2 V(x) = 3/8$\\ then, the singularity is of
the  {\em limit-point} type.\\ $\bullet$  If 
$\displaystyle  \lim_{x \rightarrow
x_0} \sup \, (x-x_0)^2 V(x) <  3/8$,\\ it is of the {\em limit-circle}
type.
\end{claim}

Indeed,  as  $\lim_{x \rightarrow  x_0}  V(x)  =  + \infty$,  applying
theorem  4  of  \cite{Fulton}  shows  immediately  that  the  type  of
singularity   of  the  branch   located  at   $x_0$  is   always  {\em
nonoscillatory}.   Whether it  is  of the  {\em  limit-point} or  {\em
limit-circle} type relies on theorem 5 of the same paper.

\subsection{Family I and Family II potentials}

Now, we investigate the spectrum of real-analytic 
potentials belonging to Family I or II numerically and compare it to the WKB 
predictions. We select two specific types of potentials for this study
with a frequency set to unity. 

For potentials of type I, we consider \eqref{VStill} with $\alpha=1/2$
($\alpha<1$  implies  that  the  potential  is nonsingular) and 
we denote it 
\begin{equation} \label{VI}
V_{I}(\beta;x) = \frac{2}{\beta^2}  \left(  (\beta
 x+1)-\sqrt{\frac{\beta x}{2}(\beta x+2)+1}\right)^2\, .
\end{equation}
Of course, some quantitative differences are to be expected according to 
the value of $\alpha$ but, qualitatively, the features we are going to discuss 
hereafter remain unchanged.

For potentials of type II, we select \eqref{Vxi3alpha1} and denote it
\begin{equation} \label{VII}
V_{I\!I}(\beta;x) = \frac{1}{8\beta^2}     \left[\ln      \left(
  \frac{\sqrt{1+8e^{3\beta x}}-1}{2} \right)\right]^2 \, .
\end{equation}
Again, we could have chosen \eqref{Vxi2alpha1} instead without 
consequences regarding the qualitative behaviour of the spectrum 
as a function of the parameter $\beta$.

\subsubsection{Small $\beta$ regime}

As already noted on the specific example of the isotonic potential
(see section \ref{sssWKBQCFI}), the WKB method is expected to provide
the best results in the limit where the parameter $\beta$ is small. In this
limit, the potential is close to being harmonic. We verify this
statement by comparing the exact numerical corrections to the harmonic levels 
to those obtained from the semiclassical quantisation condition. 
For $V_{I}(\beta;x)$, we 
use the analytical expression \eqref{Quantbol} whereas for 
$V_{I\!I}(\beta;x)$, 
we resort to a numerical evaluation of $I_2(E)$ and $I_4(E)$. Within 
this section, $\beta=1/2$. 

We  define the  exact
(numerical) correction to the harmonic levels by
\begin{equation} \label{correxact}
\varepsilon_n = E_n-(n+\frac{1}{2})
\end{equation}
and, according to \eqref{Quantcond2} and \eqref{WKBscaling}, 
its (continuous) fourth order semiclassical approximation by
\begin{equation} \label{corrWKBgen}
\varepsilon_{{}_{\rm WKB}}(E) = 
-\beta^2 I_2(\beta^2 E) -\beta^6 I_4(\beta^2 E)\, . 
\end{equation}
In this last expression, $I_2(E)$ and $I_4(E)$ are given by \eqref{I2v} 
and \eqref{I4v}, respectively, and are evaluated at $\omega=1$ for the 
functions $S(X)$ defined in \eqref{SStill} (type I) and 
\eqref{SDor} (type II).
For $V_{I}(\beta;x)$, \eqref{Quantbol} yields
\begin{equation} \label{corrWKBbol}
\varepsilon_{{}_{\rm
  WKB}}(E)=  \frac{1}{2^5}\frac{\beta^2}{Q^{5/2}}-
  \frac{\beta^6}{2^{12}}\frac{P}{Q^{15/2}} \, .
\end{equation}
where $P$ and $Q$, defined in \eqref{PandQ}, are taken at $\alpha=1/2$ and 
$\omega=1$.

\begin{figure}
\begin{center}
\scalebox{0.7}{\input{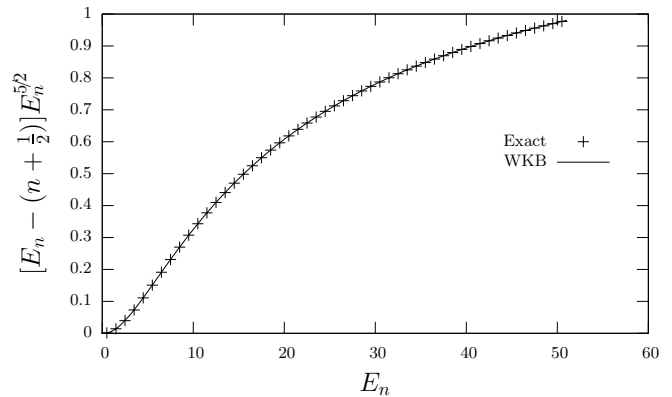}}
\caption{Comparison of the exact and semiclassical corrections 
to the harmonic spectrum for the  first 50 energy
  levels of $V_{I}(1/2;x)$ (see \eqref{VI}). Energy
  differences  $\varepsilon_n$ and $\varepsilon_{{}_{\rm WKB}}(E)$ 
(see \eqref{correxact} and \eqref{corrWKBbol}) are multiplied by $E^{5/2}$.
  Cross symbols are the numerical results and the solid line indicates
  the WKB results.}\label{Bola5m1b5m1}
\end{center}
\end{figure}

\begin{figure}
\begin{center}
\scalebox{0.7}{\input{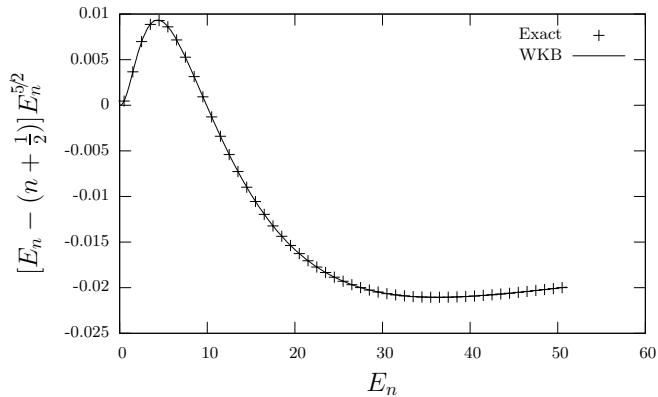}}
\caption{Comparison of the exact and semiclassical corrections 
to the harmonic spectrum for the  first 50 energy
  levels of $V_{I\!I}(1/2;x)$ (see \eqref{VII}) 
multiplied by $E^{5/2}$ as in Fig. 
\ref{Bola5m1b5m1}. Cross symbols are the numerical results and the solid
  line, the result of the numerical evaluation of \eqref{corrWKBgen}.} 
\label{Tanhthirdbeta5m1}
\end{center}
\end{figure}

The results  are presented in Figures \ref{Bola5m1b5m1} and 
\ref{Tanhthirdbeta5m1}. For the sake of visibility, in both cases, 
the correction  $\varepsilon$  is  multiplied by  the
inverse  of  its asymptotic  decay,  $E^{5/2}$  (see \eqref{I2bol} and
\eqref{asymptQCxi23}). Cross symbols represent the exact numerical
results and the solid line, the fourth-order WKB results. As we can
see, the fit is virtually perfect for the two spectra. 
In the limit where $\beta$ is small, WKB provides corrections which 
are very accurate even in the lowest part of the spectrum. And this is so 
even though, in both cases, $E^{5/2} \varepsilon_{{}_{\rm WKB}}(E)$ 
is far from reaching its asymptotic values given by
$\sqrt{2}$ (type I) and $-2M_{2,1}\simeq -1.261$ (type II), respectively. 
On these precise examples, we see 
that WKB carry much more information than a simple
high-energy asymptotic expansion. 


\subsubsection{Large $\beta$ regime}

As shown in claim \ref{asymptbehavV}, in the limit of large $\beta$, an 
isochronous potential 
asymptotically  converges  towards  a split-harmonic  oscillator. 
Thus, we expect the corresponding quantum
spectra to be close. For this reason, an analysis of 
the spectrum of the  split-harmonic potential has been done in appendix  
\ref{App2} where a high-energy asymptotic
expression for its quantum levels is derived.
Using these results, we now compare the spectrum of potentials 
$V_{I}(50;x)$ and $V_{I\!I}(30;x)$
to the spectrum of their respective asymptotic
split-harmonic potentials ($\beta \rightarrow \infty$).

\begin{figure}
\begin{center}
\scalebox{0.7}{\input{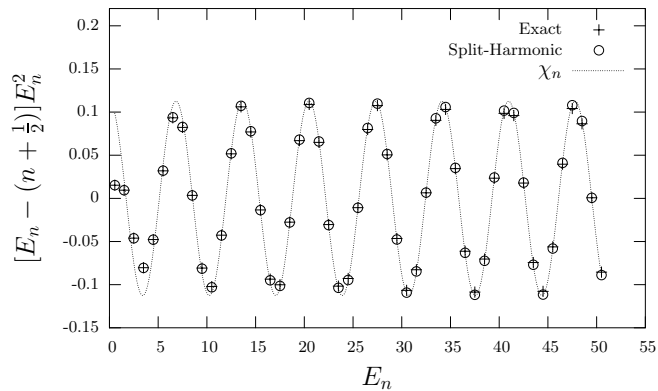}}
\caption{Comparison of the exact and split-harmonic corrections to
  the harmonic spectrum for the first 50 energy
  levels of $V_{I}(50;x)$ (see \eqref{VI}). Notice that, contrary to Fig. 
  \ref{Bola5m1b5m1}, corrections are now multiplied by $E^2$. Cross
  symbols  are  the  numerical  results, circles are  the  corresponding 
  split-harmonic
  correction and the dotted line is the analytical approximation $\chi_n$ 
  obtained in \eqref{epsSH} (last expression).}
\label{Bola5m1b5e1}
\end{center}
\end{figure}
\begin{figure}
\begin{center}
\scalebox{0.7}{\input{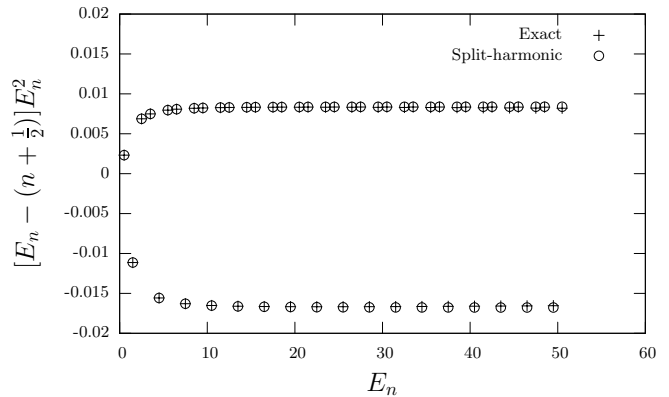}}
\caption{Comparison of the exact and split-harmonic corrections to
  the harmonic spectrum for the first 50 energy
  levels of $V_{I\!I}(30;x)$ (see \eqref{VII}). Like in Fig. 
  \ref{Bola5m1b5e1}, corrections are multiplied by $E^2$. Cross
  symbols  are  the  numerical  results and circles, the corresponding 
  split-harmonic correction.}\label{Tanhthirdbeta3e1}
\end{center}
\end{figure}

According  to  \eqref{Splitharm}, when $\beta \rightarrow \infty$, 
$V_{I}(\beta;x)$
converges to a split-harmonic oscillator with left and right
frequencies whose ratio is given by
$\rho_{{}_I}=(1-\sqrt{\alpha})/(1+\sqrt{\alpha})=3-2\sqrt{2}$.  
Regarding $V_{I\!I}(\beta;x)$, this ratio becomes $\rho_{{}_{I\!I}}=1/2$ 
(see end of section \ref{FamilyII}).
 
Again, for the sake
of clarity, we  have multiplied \eqref{correxact} by the inverse
of its  {\em split-harmonic} asymptotic decay. Notice  that the latter
is different from the WKB  decay. Indeed, as shown in \eqref{ESH}, the
{\em  split-harmonic  correction}  to  the harmonic levels  vanishes  like
$E^{-2}$ as  $E$ increases whereas  the {\em WKB  correction} decays
like $E^{-5/2}$ instead.

In Figure  \ref{Bola5m1b5e1},   even  though
$\varepsilon_n$  behaves  erratically, in contrast to
the regularity  observed in  Figure \ref{Bola5m1b5m1},
the exact numerical  results for $V_{I}(50;x)$ 
are very  well approximated by
the  split-harmonic corrections  for low-energy  levels.  The apparent
complex  behaviour  of   $\varepsilon_n$  is  easily  understood  from
expressions \eqref{ESH} and \eqref{epsSH} of appendix \ref{App2} which
show that,  once multiplied by  $E_n^2$, the correction ($\chi_n$) 
is  a periodic function of $E_n$ (or $n+1/2$) sampled at an incommensurate 
frequency. The dotted line of Figure \ref{Bola5m1b5e1}, which is a continuous 
version of $\chi_n$ (last expression of \eqref{epsSH}), 
shows how accurate this asymptotic result is.

In Figure \ref{Tanhthirdbeta3e1}, we have reported the results obtained 
for $V_{I\!I}(30;x)$. For such a large value of $\beta$, the correction 
to the harmonic levels is again very close to the split-harmonic correction.
But this time, due to the rational value of $\rho_{{}_{I\!I}}=1/2$, 
\eqref{epsSH} shows that $E_n^2\varepsilon_n \sim \chi_n$ takes on
two values only, $\frac{3^3}{2^{10}\pi}$ and $-\frac{3^3}{2^9\pi}$, 
in perfect agreement with those observed in Figure 
\ref{Tanhthirdbeta3e1}.

Yet, both Figure
\ref{Bola5m1b5e1} and Figure \ref{Tanhthirdbeta3e1} 
indicate that the exact correction $\varepsilon_n$ and
its asymptotic approximation progressively split up as the energy increases. 
This is what
is expected if the WKB analysis becomes exact as $E \rightarrow
\infty$. Then, a crossover  should exist between  the  two different
power-law decays, $E^{-2}$, which is valid  at low energy and $E^{-5/2}$, 
which is valid at  high energy.  
This is  difficult to  observe for  large  values of
$\beta$, however, because it  would require a very  high numerical
accuracy  when going  up the  spectrum. Instead,  we try  
 to  observe this transition  for an intermediate value  of the
parameter $\beta$ in  the next section.

\subsubsection{Intermediate $\beta$ regime}

\begin{figure}
\begin{center}
\scalebox{0.7}{\input{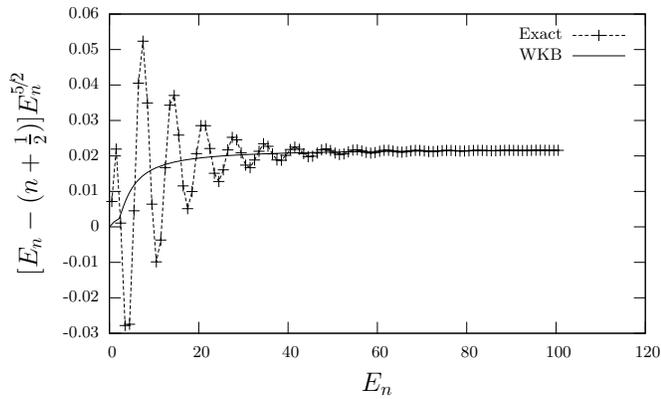}}
\caption{Comparison of the exact and semiclassical corrections 
to the harmonic spectrum for the  first 100 energy
  levels of $V_{I}(2;x)$ (see \eqref{VI}). Energy
  differences  $\varepsilon_n$ and $\varepsilon_{{}_{\rm WKB}}(E)$ 
(see \eqref{correxact} and \eqref{corrWKBbol}) are multiplied by $E^{5/2}$.
  Cross symbols linked by a dashed line 
are the numerical results and the solid line indicates the WKB results.}
\label{Bola5m1b2}
\end{center}
\end{figure}
 
\begin{figure}
\begin{center}
\scalebox{0.7}{\input{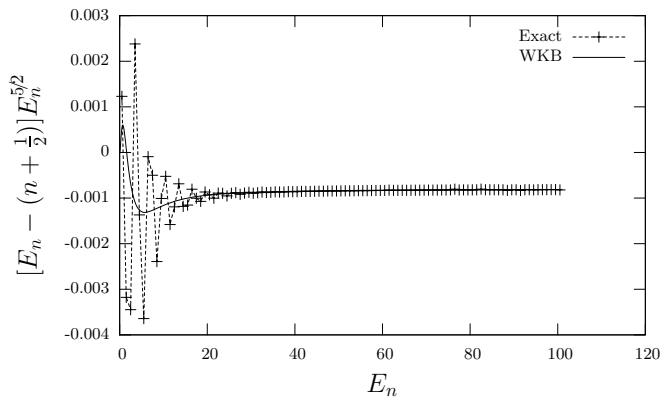}}
\caption{Comparison of the exact and semiclassical corrections 
to the harmonic spectrum for the  first 100 energy
  levels of $V_{I\!I}(\sqrt[3]{2};x)$ (see \eqref{VII}) 
multiplied by $E^{5/2}$ as in Fig. 
\ref{Bola5m1b2}. Cross symbols linked by a dashed line 
are the numerical results and the solid
line, the result of the numerical evaluation of \eqref{corrWKBgen}.}
\label{Tanhthirdbetacbrt2}
\end{center}
\end{figure}

In this last section, we select the two potentials $V_{I}(2;x)$ and 
$V_{I\!I}(\sqrt[3]{2};x)$. The  results are  shown   in 
Figure \ref{Bola5m1b2} and Figure \ref{Tanhthirdbetacbrt2}, respectively.   The
correction $\varepsilon_n$  is again multiplied  by the  inverse of its
asymptotic  WKB  decay,  $E_n^{5/2}$. In both cases, we observe some 
oscillations at  low
energy which are clearly reminiscent of those produced by a split-harmonic
oscillator. It can be checked that they are almost of the same frequency
as those observed in Figures \ref{Bola5m1b5e1} and \ref{Tanhthirdbeta3e1}. 
But at higher energy, they  are damped out
and converge towards the asymptotic behaviour predicted by the 
semiclassical approach. Indeed, from \eqref{corrWKBbol} and 
\eqref{asymptQCxi23}, we easily obtain  
$E^{5/2}\varepsilon_{{}_{\rm WKB}} \rightarrow \sqrt{2}/64 $ for $V_{I}(2;x)$
and 
$E^{5/2}\varepsilon_{{}_{\rm WKB}} \rightarrow -M_{2,1}/8$ 
  for $V_{I\!I}(\sqrt[3]{2};x)$. Hence, the asymptotic values 
$2.21 \times 10^{-2}$ and $-7.88\times 10^{-4}$ oberved in Figures 
\ref{Bola5m1b2} and \ref{Tanhthirdbetacbrt2}, respectively.

\section{Discussion and conclusion} \label{Discuss}  

In this paper, we have provided a quantitative way to analyse the 
spectrum of analytic isochronous potentials. As is already known,
 the spectrum of an isochronous potential 
is generally not strictly regularly spaced, 
in contrast to the harmonic one. Therefore, a method is required
to obtain the possible corrections to the harmonic equispacing of its
energy levels.

We have shown that, a WKB analysis going beyond the semiclassical EBK 
quantisation is precise enough to successfully account for these small
differences. It provides the right asymptotic behaviour of these corrections
at high energy, and in certain cases, is even able to describe them 
accurately from the ground-state.

Moreover, we have seen that, for any isochronous potential $V(x)$, there
exists a simple scaling transformation, namely 
$\tilde{V}(\beta;x)=V(\beta x)/\beta^2$, 
which preserves its isochronism as well 
as its frequency. Under certain regularity conditions, the one-parameter 
family $\tilde{V}(\beta;x)$ interpolates continuously between the harmonic 
oscillator ($\beta \rightarrow 0$), and a split-harmonic
oscillator ($\beta \rightarrow \infty$). General spectral features,  
with respect to the parameter $\beta$, have been sketched
on several examples. 

When $\beta$ is small, that is, when $\tilde{V}(\beta;x)$ 
is close to a harmonic oscillator, a higher-order quantisation condition 
derived from the WKB series accurately describes the entire spectrum, even
its lowest part. When $\beta$ is large, $\tilde{V}(\beta;x)$
is close to a split-harmonic oscillator and the lowest part of its spectrum 
essentially reproduces the split-harmonic features. Nevertheless, 
at high energy, the spectrum becomes asymptotically described by a 
higher-order WKB quantisation  condition again, which significantly differs
from the exact split-harmonic spectrum. 

Corrections to the harmonic spectrum, 
that is, $\varepsilon_n=E_n-(n+\frac{1}{2})\hbar \omega$, have been shown 
to scale like $E_n^{-2}$ at high energy for a split-harmonic oscillator. 
Those generated by a higher-order WKB method, however, scale generically like 
$E_n^{-5/2}$. Assuming that the latter ultimately 
represent the right high-energy 
asymptotic behaviour of the spectrum, this leads to the conclusion that a 
crossover between these two trends exists. 
Although its precise location within the spectrum is difficult 
to explore numerically, the spectrum of $\tilde{V}(\beta;x)$ for 
intermediate values of the parameter $\beta$ supports evidence of 
such a transition. 

Notice finally that, in the examples treated in the last section, the first 
term of the WKB series going beyond the EBK quantisation, namely $I_2$, 
dictates the leading order of the asymptotic decay of the correction 
$\varepsilon_n$ as $E_n \rightarrow \infty$. Moreover, in these examples, 
$\varepsilon_n \sim E_n^{-5/2}$. We would like to point out
that this needs not always be the case. Indeed,
 starting from expression \eqref{I2v} for $I_2(E)$ and using the
properties of Abel transforms \cite{Bleistein}, 
\cite{Gorenflo}, its possible to invert the problem and to calculate 
the general expression of the function $S(X)$ corresponding to a 
prescribed function $I_2(E)$ (see appendix \ref{App3}).

Using this result, we can choose $I_2(E)$ (and 
deduce the corresponding analytic isochronous potential 
via the function $S(X)$) such that, its asymptotic decay 
is faster than the asymptotic decay of $I_4(E)$. An example given in 
appendix \ref{App3} proves that it is possible to find 
analytic potentials such that $I_2(E)\sim E^{-9/2}$ and $I_4(E)\sim E^{-7/2}$
as $E \rightarrow \infty$. The asymptotic leading order is then given by
$I_4$ in this case and different from the generic $E^{-5/2}$ decay.   
This precise example indicates how difficult it is to draw some 
general conclusions regarding the behaviour of the WKB series with 
respect to the energy.

Our last remark concerns the class of isochronous potentials with a
strictly equispaced (harmonic) spectrum. We have met such a one-parameter
family in section \ref{sssWKBQCFI} called {\em isotonic} or 
{\em radial-harmonic} oscillator. The interesting question is whether 
this family is the only one to be both classical and quantum ``harmonic''.
Without this being a proof, we would like to mention the following 
interesting result: 
\begin{claim}
The most general family of analytic isochronous potentials, such that 
all the terms of the WKB series, $I_{2n}(E),\ n \geq 1$, 
defined in \eqref{I2k}, 
are constant (energy-independent), 
is the family of the isotonic oscillator with potential
$$V(\beta;x)=\frac{\omega^2}{8\beta^2}  \left(  \beta  x+1-\frac{1}{\beta
 x+1}\right)^2 \ \ \ x > -\frac{1}{\beta}\ .$$
\end{claim} 

\begin{proof}
As already proved in \cite{Roman00,Barclay94}, 
all the terms $I_{2n}(E),\ n \geq 1$ 
of the isotonic potential are constant and 
once summed, the WKB series leads to the exact quantisation 
condition for this potential. To prove that this family is the most general,
we use the result of appendix \ref{App3}, which shows that requiring $I_2(E)$
to be constant implies that $S(X)$ corresponds to the function 
of the isotonic oscillator.
\end{proof}  

Obviously, the fact that the $I_{2n}$'s,
$\ n \geq 1$, are constant ensures that the spectrum is strictly equispaced. 
Indeed, knowing that 
$I_0(E)$ is proportional to $E$, because the potential is isochronous and 
that all other terms ($I_1$ included) are constant, immediately leads 
to a quantisation condition of the form $E = \alpha(n+\mu)$, 
$(\alpha,\mu) \in \R^2$, and hence,
to a regular spacing between consecutive energy levels. 
   
Unfortunately, the claim above is no proof that the isotonic
family is the only one for which 
all terms of the WKB series add up to an energy-independent expression. 
For example, we 
could think of an overall cancelation of the energy-dependent terms
generated by the $I_{2n}(E)$'s. Moreover, although the last condition 
is quite a natural way to ensure that the spectrum is strictly equispaced, 
the function $\sum_{n=1}^{\infty}I_{2n}(E)$  
could be {\em energy-dependent} and the quantisation condition 
$\sum_{n=0}^{\infty}I_{2n}(E)= (n+\frac{1}{2})\hbar$ still have 
an equispaced spectrum for solution. Although they seem quite unlikely to us,
these possibilities cannot be excluded and the question of the class of 
potentials both classical and quantum harmonic remains open.\\

{\em Acknowledgements}
I am grateful to S. Flach for having stimulated 
my interest in isochronous potentials, to R.S. MacKay for fruitul 
exchanges on the subject and to A. Kalinowski for her  
useful comments and suggestions regarding this manuscript.

\appendix

\section{Scaling of the WKB terms} \label{App1}

Here, we show the transformation  affecting each term of the WKB series
as      the       potential      $V(x)$      is       rescaled      to
$\tilde{V}(x)=(\gamma/\beta)^2V(\beta  x)$. First we consider  the
Schr\"odinger equation for $\tilde{V}(x)$,
\begin{equation} \label{Schrod1}
-\frac{\hbar^2}{2}\frac{d^2\tilde{\psi}(x)}{dx^2}+    \tilde{V}(    x)
\tilde{\psi}(x) = \tilde{E} \tilde{\psi}(x)\, .
\end{equation}
According to \eqref{Quantcond2}, its WKB quantisation condition reads
\begin{equation} \label{Quant1}
 \sum_{k=0}^{\infty}     \tilde{I}_{2k}    (\tilde{E})     =    \left(
 n+\frac{1}{2}\right) \, \hbar \, ,\ n \in \N
\end{equation}
Given    equation    \eqref{I2k},    $\tilde{I}_{2k}(\tilde{E})$    is
proportional to $\hbar^{2k}$ and we will explicitely write it
\begin{equation} 
 \tilde{I}_{2k}   (\tilde{E})   =   \hbar^{2k}   \tilde{\cal   I}_{2k}
 (\tilde{E})\, .
\end{equation} 
Let  $y=\beta x$,  $\psi(y)=\tilde{\psi}(x)$.  Eq. \eqref{Schrod1}  is
immediately transformed to
\begin{equation} \label{Schrod2}
-\frac{\hbar^2}{2}\left(\frac{\beta^2}{\gamma}\right)^2\frac{d^2\psi(y)}{dy^2}+
V(y) \psi(y) = E \psi(y)
\end{equation}
where
\begin{equation} \label{EtildeE}
E = \left(\frac{\beta}{\gamma}\right)^2 \tilde{E}\, .
\end{equation}
Equation \eqref{Schrod2} is a  Schr\"odinger equation in the potential
$V(x)$  with  an  effective   Planck's  constant  $\hbar_{\rm  eff}  =
\frac{\beta^2\hbar}{\gamma}$. Therefore, its quantisation condition is
\begin{equation} \label{Quant2}
 \sum_{k=0}^{\infty}      \left(\frac{\beta^2\hbar}{\gamma}\right)^{2k}
 {\cal      I}_{2k}     (E)     =      \left(     n+\frac{1}{2}\right)
 \frac{\beta^2\hbar}{\gamma} \, \, ,\ n \in \N
\end{equation}
Given that  \eqref{Schrod1} and \eqref{Schrod2}  are one and  the same
equation  their   quantisation  conditions  are   the  same.  Dividing
\eqref{Quant2} by $\frac{\beta^2}{\gamma}$  and identifying the series
term by term to \eqref{Quant1}, we obtain the desired scaling
\begin{equation} \label{I2nscaling}
\tilde{I}_{2k}(E)     =     \left(\frac{\beta^2}{\gamma}\right)^{2k-1}
I_{2k}\left(\frac{\beta^2 E}{\gamma^2 }\right) \, .
\end{equation}

Another  way to  proceed is  to consider  the explicit  expression for
$I_{2n}(E)$ given  in \cite{Robnik00}. According  to Eq. (44)  of this
reference,   for   $m   \geq    1$,   $I_{2m}(E)$   can be written as
\footnote{Differences in the coefficients of this formula when compared
to Eq.  (44) of \cite{Robnik00} are  due to a different  choice in the
normalisation of the initial Schr\"odinger equation.}
\begin{equation} \label{Irob}
I_{2m}(E)        =        -\frac{\sqrt{2}}{\pi}       \sum_{L(\nu)=2m}
\frac{2^{|\nu|}J_{\nu}(E)} {(2m-3+2|\nu|)!!}
\end{equation}
where
\begin{equation}
J_{\nu}(E)   =   \frac{\partial^{m-1+|\nu|}}{\partial   E^{m-1+|\nu|}}
\int_{x_-(E)}^{x_+(E)} \frac{U_{\nu}V^{(\nu)}(x)}{\sqrt{E-V(x)}}\, dx
\end{equation}
and   where  $\nu=(\nu_1,\nu_2,\dots,\nu_{2m})$,  $\nu_j   \in  \Z^+$,
$L(\nu)=\sum_{j=1}^{2m}             j            \nu_j$            and
$|\nu|=\sum_{j=1}^{2m}\nu_j$. Moreover,
\begin{equation}
V^{(\nu)}(x)                                                          =
\prod_{j=1}^{2m}\left(\frac{d^jV}{dx^j}(x)\right)^{\nu_j}\, .
\end{equation} 
Coefficients $U_{\nu}$  are defined by a reccurrence  relation but are
not given here since they are not needed in deriving the result.

Let us evaluate $\tilde{I}_{2m}(E)$ which is related to the potential
$\tilde{V}(x)=(\gamma/\beta)^2V(\beta x)$. We first see that
\begin{equation}
\tilde{V}^{(\nu)}(x)                                                  =
\left(\frac{\gamma}{\beta}\right)^{2|\nu|}\beta^{L(\nu)}V^{(\nu)}(\beta
x) \, .
\end{equation} 
Taking  this  result  and  \eqref{scaledturnpoints} into  account,  we
obtain
\begin{equation}
\tilde{J}_{(\nu)}(E)                                                  =
\frac{1}{\gamma}\left(\frac{\beta}{\gamma}\right)^{2(m-1)}\beta^{L(\nu)}
J_{(\nu)}\left(\frac{\beta^2E}{\gamma^2}\right)
\end{equation} 
Given that  in Eq. \eqref{Irob}, the  sum is restricted  to terms such
that $L(\nu)=2m$, we finally reobtain \eqref{I2nscaling}.

\section{Spectrum of the split-harmonic potential} \label{App2}

The  {\em  split-harmonic} potential  is  made  of two  half-parabolic
arches connected  in $x=0$  with different frequencies  $\omega_l$ and
$\omega_r$ to the left and  to the right respectively. It is an
isochronous  potential       with         frequency
$\omega=2\omega_l\omega_r/(\omega_l+\omega_r)$.  It  turns out that an
exact expression for its quantisation condition is available and reads
(see for example \cite{Still89})
\begin{equation} \label{QCSH}
\frac{\sqrt{\rho}}                        {\Gamma\left(\frac{3}{4}-\rho
x\right)\Gamma\left(\frac{1}{4}-x\right)}+
\frac{1}{\Gamma\left(\frac{1}{4}-\rho
x\right)\Gamma\left(\frac{3}{4}-x\right)} = 0
\end{equation} 
In this  expression $\rho=\omega_l/\omega_r$ is the  ratio between the
left and right frequencies and without loss of generality, we restrict
its values to the range  $[0,1]$ (values within the range $[1,\infty]$
give  the  same  spectrum because they amount  to  flipping  the
potential around  its vertical axis).  Finally, $x=(\nu+1/2)/(1+\rho)$
and  $E_{\nu}=(\nu+\frac{1}{2})\omega$.  Thus,  once  \eqref{QCSH}  is
solved for $x$, we know the energy levels $E_{\nu}$.

There are  two limiting cases corresponding to the
  {\em  harmonic}  potential ($\rho=1$)  and  the {\em  half-harmonic}
  potential $(\rho=0)$  that can be treated  exactly. As $1/\Gamma(z)$
  is  an  entire function  in  $\C$  which  vanishes at  all  negative
  integers, we obtain:
\begin{itemize}
\item For $\rho=1$,  $x=3/4-n$ or $x=1/4-n$, $n \in  \N$. Thus, $\nu=n
  \in \N$ and $E_n=(n+\frac{1}{2})\omega$ which is indeed the spectrum
  of the harmonic oscillator.
\item  For $\rho=0$,  $x=3/4-n$,  $n \in  \N$.  Thus, $\nu=n+1/4$  and
$E_n=(n+\frac{3}{4})\omega$ which  is also kown to be  the spectrum of
the harmonic oscillator on the half-line.
\end{itemize}     
In these  two limiting cases,  energy levels  are exactly
equidistant within  the spectrum.  The only noticeable  change regards
the Maslov index of the half-harmonic oscillator which is $3/4$ due to
the presence of a ``wall-type''  singularity in $x=0$ (see for example
\cite{Friedrich96}).

However, for $\rho>0$, no such singularity occurs. We  expect the
spectrum  to  be  asymptotically  given by  the  semiclassical  energy
levels,  that is  $E_n \sim  (n+\frac{1}{2})\omega$ as  $n \rightarrow
\infty$.   As  it  has  already  been shown \cite{Still89},  
the  spectrum  of  the  split  harmonic  potential  is  not
regularly  spaced  for  $0<\rho<1$.  Approximate expressions  for  the
energy levels are given in  \cite{Still89} when $\rho \sim  1$ or 
$\rho \sim 0$  \footnote{These authors use the
variable     $\xi$    related     to    $\rho$ by 
$\rho=(1-\xi)/(1+\xi)$.}. Explicit  corrections are given  for the first 
five energy  levels when $\rho  \sim 1$ and  it is noticed  that {\em
``The  striking feature  of these  results is  that they  alternate in
sign''}.

To  explain  this interesting  phenomenon,  we  perform an  asymptotic
expansion of  the energy levels  $E_n$ for $\rho>0$ as  $n \rightarrow
\infty$.  Starting from \eqref{QCSH} and after some algebra, we obtain
\begin{equation} \label{ESH}
E_n                      \sim                     \left[n+\frac{1}{2}+
\frac{\chi_n}{\left(n+\frac{1}{2}\right)^2}\right]\, \omega\, ,
\ \ \ \rho n \rightarrow \infty
\end{equation} 
where
\begin{eqnarray} 
\chi_n & = & -\frac{(1+\rho)^3(1-\rho)}{128\pi        \rho^2}
\cos\left[\frac{2\pi}{1+\rho}\left(n+\frac{1}{2}\right)\right], \nonumber \\
& = & \frac{(1+\rho)^3(1-\rho)}{128\pi        \rho^2}
\cos\left[\frac{2\pi \rho}{1+\rho}\left(n+\frac{1}{2}\right)\right]\, .
\label{epsSH}
\end{eqnarray} 
Clearly, \eqref{ESH} indicates  that  $E_n$
converges asymptotically to the harmonic levels for all $\rho>0$. This
could seem  surprising since we know  that they don't  for $\rho=0$ (they
are equal  to $(n+\frac{3}{4})\omega$ instead). In  this respect, this
limit is clearly singular.  Hence, the requirement $\rho n \rightarrow
\infty$ for \eqref{ESH}.
\begin{figure}
\begin{center}
\scalebox{0.7}{\input{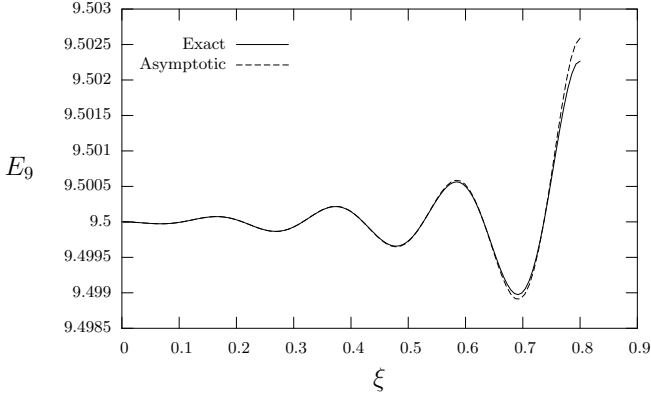}}
\caption{Parametric evolution of the ninth energy level, $E_9$, 
of a split-harmonic oscillator with frequency ratio $\rho$ with respect to
the variable $\xi=(1-\rho)/(1+\rho)$ ($\omega=1$).
The solid line indicates the exact numerical 
result obtained from \eqref{QCSH} and the dashed line 
indicates the asymptotic result
\eqref{ESH} [to be compared with Figure 3 of
  reference \cite{Still89}.]}
\label{ES9xi}
\end{center}
\end{figure}

The second remark is that, compared  to {\em Family I} and {\em Family
II}  potentials  whose levels  converge  like  $E^{-5/2}$ towards  the
harmonic ones,  the levels  of the split-harmonic  oscillator converge
like $E^{-2}$  instead.  For  isochronous potentials with  large 
parameter  $\beta$ (that is, close  to  a split-harmonic  oscillator),  we
expect a  crossover between an asymptotic $E^{-2}$  convergence law in
the low/medium part  of the spectrum and a  $E^{-5/2}$ convergence law
in the large energy limit.

Finally, eq. \eqref{epsSH} explains the oscillatory  behaviour of the
split-harmonic levels noticed by F.H Stillinger and D.K Stillinger 
in \cite{Still89} as both $n$ and
$\rho$ vary since they are involved in the
cosine  function of \eqref{epsSH}. In particular,  it can  be checked  
that  Figure  3 of  \cite{Still89}  is  reproduced  perfectly well  by
\eqref{epsSH} (see Fig. \ref{ES9xi}). 
And the $n/2$ oscillations of  $E_n(\rho)$ within the range $\rho \in [0,1]$
are also explained by \eqref{epsSH}.

\begin{figure}
\begin{center}
\scalebox{0.7}{\input{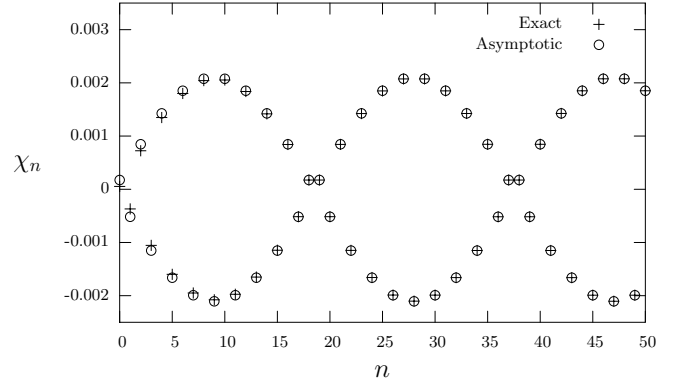}}
\caption{Corrections to the harmonic levels vs the energy level $n$, 
$\chi_n=(n+\frac{1}{2})^2[E_n-(n+\frac{1}{2})]$, for a split-harmonic potential
with  frequency ratio  $\rho=0.9$. Cross symbols
represent the exact numerical result obtained by solving \eqref{QCSH} and
circles, the asymptotic result  $\chi_n$ given  by \eqref{epsSH}.}
\label{alpha09} 
\end{center}
\end{figure}

\begin{figure}
\begin{center}
\scalebox{0.7}{\input{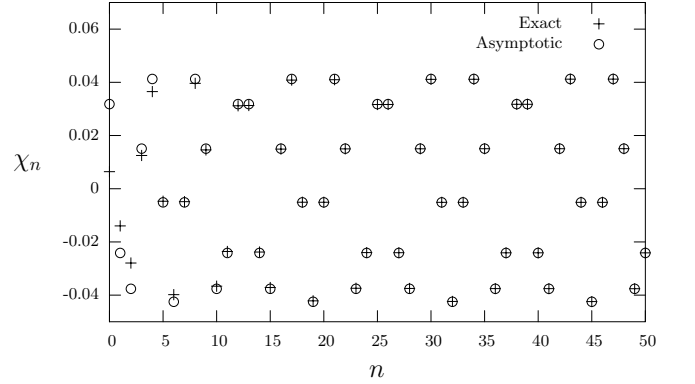}}
\caption{Same as Fig. \ref{alpha09} but for $\rho=0.3$.}
\label{alpha03} 
\end{center}
\end{figure}

In  Figs.   \ref{alpha09}  and  \ref{alpha03},  we   give  two  typical
corrections  to the  harmonic spectrum  for $\rho=0.9$  and $\rho=0.3$
respectively. 
We    define     $\chi^{\rm     exact}_n    =
(n+\frac{1}{2})^2[E^{\rm exact}_n  - (n+\frac{1}{2})]$ and  compare it
with  $\chi_n$ given  by  \eqref{epsSH} ($\omega =1$). 
Exact  values for  the
energy levels are obtained by solving \eqref{QCSH} numerically. We see
that  $\chi^{\rm  exact}_n$  and  $\chi_n$ are  in  good
agreement even for small values of $n$. But as a general trend,
this agreement degrades as $\rho \rightarrow 0$. 

\section{From $I_2(E)$ to $S(X)$.} \label{App3}

In this last appendix, we investigate the inverse problem which 
consists in recovering the potential (or at least the $S$ function) 
corresponding to a prescribed function $I_2(E)$.
Starting from expression \eqref{I2v} for $I_2(E)$, we have
\begin{equation} \label{I2vbis}
 \frac{1}{\sqrt{\pi}}
  \int\limits_{0}^{E} \!  \, \frac{g(v)}{(E-v)^{1/2}} dv= 
-\frac{12\sqrt{\pi}E^2 I_2(E)}{\hbar^2\omega}  \, .
\end{equation}
where
\begin{equation} \label{gv}
g(v) = v^{3/2}\frac{d^2 }{d
  v^2}\left[
  \frac{v}{1-S^2\left(\frac{\sqrt{2v}}{\omega}\right)}\right]\, .
\end{equation}
Now, using the properties of Abel transforms (cf. for example 
\cite{Bleistein,Gorenflo}), we invert \eqref{I2vbis} to obtain
\begin{equation} \label{gvofI2}
 g(v) = -\frac{12}{\hbar^2\omega}\frac{d}{dv}
  \int\limits_{0}^{v} \!  \, \frac{E^2 I_2(E)}{(v-E)^{1/2}} dE\, .
\end{equation}
Provided the integral on the r.h.s. of \eqref{gvofI2} can be evaluated,
we can use \eqref{gv} to calculate $S(X)$.

\subsection{First example: $I_2(E)=cst$.}

Let us assume that $I_2(E)=I_2=cst$. The r.h.s. of \eqref{gvofI2}
yields 
\begin{equation} \label{gvI2cst}
g(v) = -\frac{32I_2}{\hbar^2\omega} v^{3/2}\, .
\end{equation}  
Reinstating this expression in \eqref{gv}, we find
\begin{equation} \label{SI2cst1}
\frac{v}{1-S^2\left(\frac{\sqrt{2v}}{\omega}\right)} = 
-\frac{32I_2}{\hbar^2\omega}\left(\frac{v^2}{2}+av+b\right)
\end{equation}  
where $a$ and $b$ are two constants of integration. Since we are dealing with 
analytic potentials, $S$ is continuous in $0$ and as it is odd, $S(0)=0$.
Passing to the limit $v \rightarrow 0$ in \eqref{SI2cst1} 
shows that $b=0$. Hence,
\begin{equation} \label{SI2cst2}
1-S^2\left(\frac{\sqrt{2v}}{\omega}\right) = 
-\frac{\hbar^2\omega}{16I_2}\frac{1}{v+2a}\, .
\end{equation}
Now, the condition $S(0)=0$ determines the last constant, 
$a=-\hbar^2\omega/(32I_2)$ and 
\begin{equation} \label{SI2cst3}
S\left(\frac{\sqrt{2v}}{\omega}\right) = 
\frac{\sqrt{v}}{\sqrt{v-\frac{\hbar^2\omega}{16I_2}}}
\end{equation}
which determines $S(X)$ for $X\geq 0$ and, as $S(X)=-S(-X)$, $S(X)$
on the entire real line. Finally, the requirement $|S(X)|<1$
leads to the conclusion that $I_2$ has to be negative.
Let us write $I_2=-\hbar^2\beta^2/(8\omega)$ for some parameter 
$\beta \geq 0$, \eqref{SI2cst3} simplifies to
\begin{equation} \label{SI2cst4}
S(X) = \frac{\beta X}{\sqrt{1+\beta^2 X^2}}
\end{equation}
which is nothing but the $S$ function of an isotonic potential with 
scaling parameter $\beta$ (see \eqref{SStill}, \eqref{Isotonic} 
and \eqref{scaledxbarS}).

\subsection{Second example: 
$I_2(E)=-\frac{1}{6}\frac{\hbar^2\omega^8}{(\omega^2+2E)^{9/2}}$.}

Repeating what has been done in the previous example for the function
$I_2(E)$ chosen above yields the following $S$ function
\begin{equation} \label{SI2Em9d2}
S(X)=\frac{2X[35+42X^2+15X^4]^{1/2}}{[105+455X^2+483X^4+165X^6]^{1/2}}\, .
\end{equation}
This function is readily odd and $\forall X \in \R,\ |S(X)|<1$. The 
corresponding potential has no singularity on the real line.

For \eqref{SI2Em9d2}, the 
asymptotic behaviour of the fourth WKB
term $I_4(E)$, given by \eqref{I4v}, is proportional to $E^{-7/2}$. 
To see this, we reinstate \eqref{SI2Em9d2} into the function $G_1(X)$ 
defined in \eqref{G12} and we perform an asymptotic analysis of the 
first term of \eqref{I4v}. It turns out that,  
$g_1(v) := v^{5/2}\frac{d^3}{dv^3}\left\{G_1(\sqrt{2v}/\omega)\right\} 
\sim v^{-3/2}$ as $v \rightarrow \infty$ and the Mellin transform 
$M[g_1,1]=\int_0^{\infty}\!\! g_1(v)\, dv \neq 0$.
Thus, according to formula (4.10.25) of \cite{Bleistein}, the first 
term of \eqref{I4v} scales like $E^{-7/2}$ as $E \rightarrow \infty$.
It can be verified in the same way that the second term scales like
$E^{-9/2}$ and is negligible when compared to the first one.

\end{document}